\documentclass[conference]{IEEEtran}
\usepackage{amsmath,cite,amsfonts,amssymb,psfrag,amsthm}
\usepackage{graphicx}
\usepackage{epstopdf}
\usepackage{rotating}
\usepackage{dsfont}
\usepackage{color}
\usepackage{tikz}
\usetikzlibrary{plotmarks}
\usepackage{pgfplots}
\usetikzlibrary{calc}
\usetikzlibrary{shapes,arrows}
\usetikzlibrary{decorations.markings}
\usetikzlibrary{positioning}
\pgfplotsset{compat=1.10}
\usetikzlibrary{calc}
\usetikzlibrary{shapes,arrows}
\usetikzlibrary{decorations.markings}

\usepackage[utf8]{inputenc}
\usepackage{authblk}

\DeclareFontFamily{U}{mathx}{\hyphenchar\font45}
\DeclareFontShape{U}{mathx}{m}{n}{<-> mathx10}{}
\DeclareSymbolFont{mathx}{U}{mathx}{m}{n}
\DeclareMathAccent{\widebar}{0}{mathx}{"73}

\makeatletter
\newtheorem*{rep@theorem}{\rep@title}
\newcommand{\newreptheorem}[2]{%
	\newenvironment{rep#1}[1]{%
		\def\rep@title{\Cref{##1}}%
		\begin{rep@theorem}}%
		{\end{rep@theorem}}}
\newcommand*{\textlabel}[2]{%
	\edef\@currentlabel{#1}
	\phantomsection
	#1\label{#2}
}
\makeatother

\newtheorem{theorem}{Theorem}

\newtheorem{remark}{Remark}
\newtheorem{definition}{Definition}

\newtheorem{lemma}{Lemma}

\begin{document}
\title{Biometric and Physical Identifiers with Correlated Noise for Controllable Private Authentication}
\IEEEoverridecommandlockouts

  \author[1]{Onur G\"unl\"u}
\author[1]{Rafael F. Schaefer}
\author[2]{H. Vincent Poor}

\affil[1]{Information Theory and Applications Chair, Technische Universit\"{a}t Berlin\\ \authorcr
	\{guenlue, rafael.schaefer\}@tu-berlin.de}
\affil[2]{Department of Electrical Engineering, Princeton University\\ \authorcr
	poor@princeton.edu}

\maketitle

\begin{abstract}
	
The problem of secret-key based authentication under privacy and storage constraints on the source sequence is considered. The identifier measurement channels during authentication are assumed to be controllable via a cost-constrained action sequence. Single-letter inner and outer bounds for the key-leakage-storage-cost regions are derived for a generalization of a classic two-terminal key agreement model with an eavesdropper that observes a sequence that is correlated with the sequences observed by the legitimate terminals. The additions to the model are that the encoder observes a noisy version of a remote source, and the noisy output and the remote source output together with an action sequence are given as inputs to the measurement channel at the decoder. Thus, correlation is introduced between the noise components on the encoder and decoder measurements. The model with a secret key generated by an encoder is extended to the randomized models, where a secret-key is embedded to the encoder. The results are relevant for several user and device authentication scenarios including physical and biometric identifiers with multiple measurements that provide diversity and multiplexing gains. To illustrate the behavior of the rate region, achievable (secret-key rate, storage-rate, cost) tuples are given for binary identifiers and measurement channels that can be represented as a mixture of binary symmetric subchannels. The gains from using an action sequence such as a large secret-key rate at a significantly small hardware cost, are illustrated to motivate the use of low-complexity transform-coding algorithms with cost-constrained actions.
\end{abstract}

\IEEEpeerreviewmaketitle
\section{Introduction} \label{sec:intro}
A traditional method for security is to store secret keys used for, e.g., device authentication in a hardware-protected non-volatile memory (NVM). Biometric identifiers such as fingerprints and physical identifiers such as random and unique oscillation frequencies of ring oscillators (ROs) are secure and cheap alternatives to key storage in an NVM. Physical unclonable functions (PUFs) are physical identifiers that are challenge response mappings such that it is easy to evaluate the response to a given challenge and hard to guess the response to a randomly chosen challenge \cite{GassendThesis}. PUFs can be used as a source of local randomness for the wiretap channel (WTC) \cite{WTC}, where the optimal coding scheme requires random sequences at the WTC encoder. Other applications of PUFs are Internet-of-Things (IoT) device security, intellectual property (IP) protection in a field programmable gate array (FPGA), and non-repudiation \cite[Chapter 1]{benimdissertation}.

We extend the source model for key agreement from \cite{AhlswedeCsiz,Maurer} to consider multiple improvements to private authentication with PUFs and biometrics. The original model considers that an encoder observes a source output to generate a key and to send public information, called \textit{helper data}, to a decoder. Key agreement at the decoder is successful when the decoder observes a noisy source output and enough amount of helper data to reconstruct the same key. The secrecy measure used is the amount of information the helper data leaks about the secret key, i.e., \textit{secrecy leakage}, which should be negligible. Similarly, \cite{IgnaTrans,LaiTrans} argue that the information leaked about the source output should be also small to keep \emph{privacy leakage} as small as possible, which cannot be made negligible for general cases. The amount of public storage should also be kept small to limit the hardware cost \cite{csiszarnarayan,bizimWZ}. 

Biometric and physical identifier outputs are noisy. Suppose we have multiple measurements of an identifier source at the encoder, which assumes that the source is hidden or remote \cite{bizimMMMMTIFS}. The source, noisy identifier, and measurement symbol strings are proposed in \cite{bizimITW} to be related by a broadcast channel (BC) with one input and two outputs to capture the effects of correlated noise in the measurements. For instance, the surrounding hardware logic is the main reason for the noise components on encoder and decoder measurements of the same RO to be correlated \cite{MerliROCorrelated}. Motivated by the use of different identifier-measurement forms, e.g., the use of multiple measurements or variations in the quality of the measurement process \cite{permuter}, we extend a private authentication model from \cite{bizimKittipongTIFS}. In this model, identifier measurements are represented by a cost-constrained action-dependent side information acquisition where an action sequence determines the measurement channel at the decoder. A high cost for an action represents, e.g., the use of a high quality measurement device that results in a small error probability. We combine the BC measurement model with the action-dependent private authentication model to consider the correlated noise on encoder and decoder measurements such that the decoder measurement channel can be adapted to the variations in the ambient temperature, supply voltage, and surrounding logic. 

Suppose the encoder generates a key from a noisy identifier output. We call this model the \emph{generated-secret (GS)} model. Similarly, for the \emph{chosen-secret (CS)} model, a chosen key and noisy identifier measurements are combined to generate the helper data. We consider also the CS model to address the cases where the encoder, e.g., a hardware manufacturer (for PUFs) or a trusted entity (for biometrics), pre-determines the secret key for practical reasons. Key agreement with correlated side information at the eavesdropper (EVE) is considered in, e.g., \cite{SecrecyviaSourcesandChannels,Blochpaper,Khisti,AminOurKeyAgreementArxiv,HimanshuShun}. This assumption is realistic for key agreement with biometric identifiers since an eavesdropper can obtain side information from, e.g., any object touched by an individual. PUF outputs are permanently changed by invasive attacks \cite{PappuThesis}, so non-invasive attacks to the devices that embody the PUF should be considered to make this model realistic for physical identifiers. We allow side information at the eavesdropper to consider both biometric and physical identifier models. Furthermore, we study independent and identically distributed (i.i.d.) source outputs and memoryless measurement channels. These models are realistic if one uses transform-coding algorithms from \cite{bizimMDPI,Transformbio} to extract almost i.i.d. sequences from PUFs or biometric identifiers. 

We derive achievable key-leakage-storage-cost regions for a decoder measurement channel with three inputs and two outputs for a strong secrecy metric. The model of the separate encoder and decoder measurements in \cite{bizimKittipongTIFS} corresponds to a physically-degraded measurement channel for a weak secrecy metric. To provide strong secrecy, there is a ``private" key assumption, e.g., in  \cite{IgnaTrans,LaiTrans,MatthieuPolar}, where they consider that the key is available to the encoder and decoder and is hidden from an eavesdropper. This assumption is unrealistic because if a private-key protection against attackers is feasible, then there is no need for key agreement with identifiers. We do not make such unrealistic assumptions to provide strong secrecy and we use the output statistics of random binning (OSRB) method from \cite{OSRBAmin}, which requires only local randomness. Our rate regions recover previous rate regions in the literature for hidden and visible sources. We establish outer bounds for conditionally less-noisy (CLN) channels defined in \cite{RoyCondLN}. The inner bounds and outer bounds for the GS models are extended to the randomized models, i.e., CS models.

This paper is organized as follows. In Section~\ref{sec:problem_setting}, we describe our models and the problem. We give achievable key-leakage-storage-cost regions for the GS and CS models in Section~\ref{sec:InnerBounds}. We define CLN channels and show outer bounds for the key-leakage-storage-cost regions in Section~\ref{sec:Outerbounds} if two CLN conditions are satisfied. We give an example in Section~\ref{sec:example} to illustrate the gains from having a larger hardware area available for public storage in combination with a cost-constrained action sequence for a practical PUF design. Achievability proofs for the inner bounds and outer bounds for CLN channels are given in Sections~\ref{sec:InnerBoundsProof} and \ref{sec:OuterBoundProofs}, respectively. Section~\ref{sec:conclusion} concludes the paper.

\begin{figure}
	\centering
	\resizebox{0.87\linewidth}{!}{
		\begin{tikzpicture}
		\node (so) at (-1.5,-3.5) [draw,rounded corners = 5pt, minimum width=1.0cm,minimum height=0.8cm, align=left] {$P_X$};
		\node (a) at (0,-0.5) [draw,rounded corners = 6pt, minimum width=3.2cm,minimum height=0.6cm, align=left] {$
			(W,S) \overset{(a)}{=} f_1(\widetilde{X}^n)$\\$W \overset{(b)}{=} f_2(\widetilde{X}^n,S)$};
		\node (c) at (5,-3.3) [draw,rounded corners = 5pt, minimum width=1.3cm,minimum height=0.6cm, align=left] {$P_{YZ|X\widetilde{X}A}$};
		\node (f) at (0,-2.25) [draw,rounded corners = 5pt, minimum width=1cm,minimum height=0.6cm, align=left] {$P_{\widetilde{X}|X}$};
		\node (b) at (5,-0.5) [draw,rounded corners = 6pt, minimum width=3.2cm,minimum height=0.9cm, align=left] {$A^n=f_{\text{a}}(W)$\\\smallskip$\hat{S} = g\left(W,Y^n\right)$};
		\node (g) at (5,-5) [draw,rounded corners = 5pt, minimum width=1cm,minimum height=0.6cm, align=left] {EVE};
		\draw[decoration={markings,mark=at position 1 with {\arrow[scale=1.5]{latex}}},
		postaction={decorate}, thick, shorten >=1.4pt] (a.east) -- (b.west) node [midway, above] {$W$};
		\node (a1) [below of = a, node distance = 3cm] {$X^n$};
		\draw[decoration={markings,mark=at position 1 with {\arrow[scale=1.5]{latex}}},
		postaction={decorate}, thick, shorten >=1.4pt] ($(c.north)+(0.3,0)$) -- ($(b.south)+(0.3,0)$) node [midway, right] {$Y^n$};
		\draw[decoration={markings,mark=at position 1 with {\arrow[scale=1.5]{latex}}},
		postaction={decorate}, thick, shorten >=1.4pt] (so.east) -- (a1.west);
		\draw[decoration={markings,mark=at position 1 with {\arrow[scale=1.5]{latex}}},
		postaction={decorate}, thick, shorten >=1.4pt] (a1.north) -- (f.south);
		\draw[decoration={markings,mark=at position 1 with {\arrow[scale=1.5]{latex}}},
		postaction={decorate}, thick, shorten >=1.4pt] (f.north) -- (a.south) node [midway, left] {$\widetilde{X}^n$};
		\draw[decoration={markings,mark=at position 1 with {\arrow[scale=1.5]{latex}}},
		postaction={decorate}, thick, shorten >=1.4pt, dashed] (a1.east) -- ($(c.west)-(0,0.2)$) node [below left] {$X^n$};
		\draw[decoration={markings,mark=at position 1 with {\arrow[scale=1.5]{latex}}},
		postaction={decorate}, thick, shorten >=1.4pt] ($(b.south)-(0.3,0)$) -- ($(c.north)-(0.3,0)$)node [midway, left] {$A^n$};
		\draw[decoration={markings,mark=at position 1 with {\arrow[scale=1.5]{latex}}},
		postaction={decorate}, thick, shorten >=1.4pt] (c.south) -- (g.north) node [midway, right] {$Z^n$};
		\node (a2) [above of = a, node distance = 1.8cm] {$S$};
		\node (b2) [above of = b, node distance = 1.8cm] {$\hat{S}$};
		\draw[decoration={markings,mark=at position 1 with {\arrow[scale=1.5]{latex}}},
		postaction={decorate}, thick, shorten >=1.4pt] (b.north) -- (b2.south);
		\draw[decoration={markings,mark=at position 1 with {\arrow[scale=1.5]{latex}}},
		postaction={decorate}, thick, shorten >=1.4pt]  ($(a2.south)+(0.3,0)$)-- ($(a.north)+(0.3,0)$) node [midway, right] {$(b)$};
		\draw[decoration={markings,mark=at position 1 with {\arrow[scale=1.5]{latex}}},
		postaction={decorate}, thick, shorten >=1.4pt]  ($(a.north)-(0.3,0)$)-- ($(a2.south)-(0.3,0)$) node [midway, left] {$(a)$};;
		\draw[decoration={markings,mark=at position 1 with {\arrow[scale=1.5]{latex}}},
		postaction={decorate}, thick, shorten >=1.4pt] ($(a.east)+(1,0)$) -- ($(a.east)+(1,-4.5)$) -- ($(a.east)+(1,-4.5)$) -- (g.west) node [above left=0.0cm and 0.5cm of g.west] {$W$};
		\draw[decoration={markings,mark=at position 1 with {\arrow[scale=1.5]{latex}}},
		postaction={decorate}, thick, shorten >=1.4pt, dashed] ($(f.north)+(0,0.2)$) -- ($(f.north)+(3.4,0.2)$) --  ($(f.north)+(3.4,-1.3)$) -- ($(c.west)+(0,0.1)$) node [above left] {$\widetilde{X}^n$};
		\end{tikzpicture}
	}
	\caption{A hidden identifier source: $(a)$ represents the GS model with the encoder $f_1(\cdot)$ and $(b)$ represents the CS model with the encoder $f_2(\cdot,\cdot)$.}\label{fig:hidden}
\end{figure}
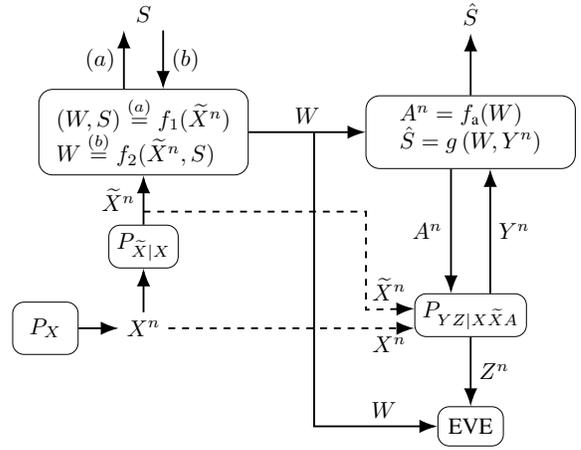

\section{Problem Definitions}\label{sec:problem_setting}
Consider the GS model in Fig.~\ref{fig:hidden}$(a)$, where a key is generated from a hidden source. The source $\mathcal{X}$, measurement $\widetilde{\mathcal{X}},\mathcal{Y}, \mathcal{Z}$, and action $\mathcal{A}$ alphabets are finite sets. The encoder observes uncontrollable noisy measurements $\widetilde{X}^n$ of the i.i.d. hidden source outputs $X^n$ through a memoryless channel $P_{\widetilde{X}|X}$. The encoder computes a secret key $S$ and public helper data $W$ as $(W,S)= f_1(\widetilde{X}^n)$. During authentication, the action encoder observes the helper data $W$ and computes an action sequence $A^n$ as $A^n = f_{\text{a}}(W)$. Then, the decoder, given $(X^n, \widetilde{X}^n, A^n)$, observes cost-constrained controllable source measurements $Y^n$ through a memoryless channel $P_{YZ|X\widetilde{X}A}$ together with the helper data $W$ and estimates the secret key as $\hat{S}=g(W,Y^n)$. The eavesdropper observes $Z^n$ as the output of the same memoryless channel in addition to the public helper data $W$. Similarly, Fig.~\ref{fig:hidden}$(b)$ shows the CS model, where a secret key $S$ that is independent of $(X^n, \widetilde{X}^n, Y^n, Z^n)$ is embedded into the helper data as $W= f_2(\widetilde{X}^n,S)$. The action encoder and the decoder for the CS model are applied in a similar way to the GS model.

\begin{definition}
	\normalfont A key-leakage-storage-cost tuple $(R_{\text{s}}, R_\ell,R_{\text{w}},C)$ is \emph{achievable} for the GS or CS model if, given $\delta\!>\!0$, there is some $n\!\geq\!1$, an encoder, and a decoder such that $R_{\text{s}}\!=\!\frac{\log|\mathcal{S}|}{n}$ and
	\begin{align}
	&\Pr[\hat{S} \neq S] \leq \delta&&\quad \quad(\text{reliability})\label{eq:reliability_cons}\\
	& I(S;W,Z^n) \leq \delta&&\quad\quad(\text{strong secrecy})\label{eq:secrecyleakage_cons}\\
	&\frac{1}{n}H(S) \geq R_{\text{s}}- \delta&&\quad\quad(\text{uniformity}) \label{eq:uniformity_cons}\\	
	&\frac{1}{n}I(X^n;W,Z^n) \leq R_\ell+\delta&&\quad\quad(\text{privacy})\label{eq:leakage_cons}\\
	&\frac{1}{n}\log\big|\mathcal{W}\big| \leq R_{\text{w}}+\delta&&\quad\quad(\text{storage})\label{eq:storage_cons}\\
	&\mathbb{E}[\Gamma(A^n)] \leq C+\delta&&\quad\quad(\text{action cost})\label{eq:cost_cons}
	\end{align}
   where we have  $\Gamma(A^n)\!=\!\frac{1}{n}\sum_{i=1}^n\Gamma(A_i)$. The \emph{key-leakage-storage-cost regions} $\mathcal{R}_{\text{gs}}$ and $\mathcal{R}_{\text{cs}}$ for the GS and CS models, respectively, are the closures of the sets of all achievable tuples for the corresponding models.\hfill $\lozenge$
\end{definition}

\section{Inner Bounds}\label{sec:InnerBounds}
We are interested in characterizing the optimal trade-off among the secret-key rate, privacy-leakage rate, storage rate, and action cost with strong secrecy for correlated noise on the encoder and decoder measurements. We give achievable rate regions for the GS and CS models in Theorem~\ref{theo:ActionBCgscs}. See Section~\ref{sec:InnerBoundsProof} for proofs. Define 
\begin{align*}
&R_{\ell,1} = I(V,X;Z|A)+I(X;A,V,Y)-I(V,X;Y|A)\\
&R_{\ell,2} =  I(V,X;Z|A,U)+I(X;A,V,Y)\nonumber\\
&\quad\qquad-I(V,X;Y|A,U)\\
&R_{\ell,3} =I(X;A,U,Z).
\end{align*}

\begin{theorem}[Inner Bounds for GS and CS Models]\label{theo:ActionBCgscs}
	An inner bound for the rate region $\mathcal{R}_{\text{gs}}$ for the GS model is the set of all tuples $(R_{\text{s}}, R_\ell,R_{\text{w}},C)$ satisfying
	\begin{align}
	&0\leq R_{\text{s}}\!\leq\! I(V;Y|A,U)- I(V;Z|A,U)\label{eq:GSRs}\\
	&R_\ell\!\geq \max\big\{R_{\ell,1},R_{\ell,2},R_{\ell,3}\big\}	\label{eq:GSRl}\\
	&R_{\text{w}}\geq I(\widetilde{X};A) + I(V;\widetilde{X}|A,Y) \label{eq:GSRw}
	\end{align}
	for some $P_XP_{\widetilde{X}|X}P_{A|\widetilde{X}}P_{YZ|X\widetilde{X}A}P_{V|\widetilde{X}A}P_{U|V}$ such that $\mathbb{E}[\Gamma(A)]\!\leq\! C$. 
	
	Similarly, an inner bound for the rate region $\mathcal{R}_{\text{cs}}$ for the CS model is the set of all tuples $(R_{\text{s}}, R_\ell,R_{\text{w}},C)$ satisfying (\ref{eq:GSRs}), (\ref{eq:GSRl}), and
	\begin{align}
	&R_{\text{w}}\geq I(\widetilde{X};A,V)-I(U;Y|A)-I(V;Z|A,U)\label{eq:CSRw}
	\end{align}
	for some $P_XP_{\widetilde{X}|X}P_{A|\widetilde{X}}P_{YZ|X\widetilde{X}A}P_{V|\widetilde{X}A}P_{U|V}$ such that $\mathbb{E}[\Gamma(A)]\!\leq\! C$. 
\end{theorem}
\begin{IEEEproof}[Proof Sketch]
	The proof for the GS model uses the OSRB method that assigns random bin indices to sequences $a^n$, $u^n$, and $v^n$ to obtain strong secrecy. Using the OSRB method consecutively, nine cases are analyzed, resulting in six different terms whose maximum is used in (\ref{eq:GSRl}). The proof for the CS model uses the key generated by using the proof for the GS model and applies a one-time padding step to the embedded secret key and the key generated by the GS model. The main effect is the increase in the storage rate as compared to the GS model by the amount equal to the bound in (\ref{eq:GSRs}). 
\end{IEEEproof}

In \cite{bizimKittipongTIFS}, separate measurements $P_{X\widetilde{X}AYZ}=P_{A|\widetilde{X}}P_{\widetilde{X}|X}P_XP_{YZ|XA}$ and a weak secrecy constraint such that (\ref{eq:secrecyleakage_cons}) is replaced with $\frac{1}{n}I(S;W,Z^n)\leq \delta$, are considered. The model in Fig.~\ref{fig:hidden} extends \cite{bizimKittipongTIFS} by considering correlation in the noise components on the encoder and decoder measurements with strong secrecy. Such a correlation is considered in \cite{bizimITW} for a model without cost-constrained action-dependent measurements at the decoder and without correlated side information $Z^n$ at the eavesdropper. Broadcast channel (BC) measurements are considered in \cite{bizimITW} to model the correlation in the noise components. Due to the causal dependence of $A^n$ on $\widetilde{X}^n$, one cannot model $\displaystyle {\widetilde{X}}^n$ as an output of the action-dependent measurement channel, so Fig.~\ref{fig:hidden} considers the encoder measurement $\widetilde{X}^n$ as an input to the measurement channel $P_{YZ|X\widetilde{X}A}$. This model is the case, e.g., if the decoder and encoder measurements are made within a coherence time, in analogy to wireless communication systems, so the encoder measurements $\widetilde{X}^n$ affect RO outputs at the decoder due to remanining temperature and current effects on digital circuits. A similar model is considered in \cite[Fig. 9]{permuter} for a source coding with side information problem without secrecy, privacy, and secret-key rate constraints.

\begin{remark}
	\normalfont The bounds in Theorem~\ref{theo:ActionBCgscs} recover the key-leakage-storage-cost regions given in \cite[Theorems 3 and 4]{bizimKittipongTIFS} for the separate-measurement model such that $P_{X\widetilde{X}AYZ}=P_{A|\widetilde{X}}P_{\widetilde{X}|X}P_XP_{YZ|XA}$ since we have
	\begin{align}
		&R_{\ell,2}\overset{(a)}{=}I(X;Z|A,U)\!+\!I(X;A,V,Y)\!-\!I(X;Y|A,U)
	\end{align}
	where $(a)$ follows for the separate-measurement model because $(U,V)-(A,X)-(Z,Y)$ form a Markov chain for this model. Similarly, the key-leakage-storage rate regions given in \cite{bizimMMMMTIFS} for measurement channels such that $P_{X\widetilde{X}Y}=P_XP_{\widetilde{X}|X}P_{Y|X}$ are recovered by the bounds in Theorem~\ref{theo:ActionBCgscs} if we choose $(Z,U,A)$ as constants. Theorem~\ref{theo:ActionBCgscs} bounds recover the key-leakage and key-leakage-storage regions for visible source models, where encoder measurements are noiseless such that $\widetilde{X}^n=X^n$, given in \cite{IgnaTrans,LaiTrans}, \cite[Theorems~1 and~2]{bizimKittipongTIFS}. 
\end{remark}


\section{Outer Bounds}\label{sec:Outerbounds}
We give outer bounds for CLN channels, defined in Definition~\ref{def:condLN}, for the model depicted in Fig.~\ref{fig:hidden}.
\begin{definition}[\hspace{1sp}\cite{RoyCondLN}]\label{def:condLN}
	\normalfont $X$ is \textit{conditionally less-noisy (CLN)} than $Z$ given $(A,Y)$ if 
	\begin{align}
		I(L;X|A,Y)\geq I(L;Z|A,Y)\label{eq:condforCLN}
	\end{align}
	holds for any random variable $L$ such that $L-(A,\widetilde{X},Y)-(X,Z)$ form a Markov chain and we denote this relation as $$(X\geq Z|A,Y).$$
\end{definition}
The set of CLN channels is shown in \cite{RoyCondLN} to be larger than the set of \textit{physically degraded} channels. We give outer bounds for the rate regions $\mathcal{R}_{\text{gs}}$ and $\mathcal{R}_{\text{cs}}$ in Theorem~\ref{theo:ActionBCGSCSOuter} when two CLN conditions are satisfied. See Section~\ref{sec:OuterBoundProofs} for proofs.

\begin{theorem}[Outer Bounds for GS and CS Models]\label{theo:ActionBCGSCSOuter}
 An outer bound for the rate region $\mathcal{R}_{\text{gs}}$ for all CLN channels such that $(X\geq Z|A,Y)$ and $(Z\geq Y|A,X)$ is the set of all tuples $(R_{\text{s}}, R_\ell,R_{\text{w}},C)$ satisfying
\begin{align}
&0\leq R_{\text{s}}\!\leq\! I(V;Y|A,U)- I(V;Z|A,U)\label{eq:GSRsOuter}\\
&R_\ell\!\geq I(X;A,V,Y) \!-\!I(X;Y|A)\!+\!I(X;Z|A) \nonumber\\
&\;\qquad+ I(U;Y|A)-I(U;Z|A)	\label{eq:GSRlOuter}\\
&R_{\text{w}}\geq I(\widetilde{X};A) + I(V;\widetilde{X}|A,Y) \label{eq:GSRwOuter}
\end{align}
such that $U-V-(A,\widetilde{X})-(A,\widetilde{X},X)-(Y,Z)$ form a Markov chain and $\mathbb{E}[\Gamma(A)]\!\leq\! C$. 

Similarly, an outer bound for the rate region $\mathcal{R}_{\text{cs}}$ for all CLN channels such that $(X\geq Z|A,Y)$ and $(Z\geq Y|A,X)$ is the set of all tuples $(R_{\text{s}}, R_\ell,R_{\text{w}},C)$ satisfying (\ref{eq:GSRsOuter}), (\ref{eq:GSRlOuter}), and
\begin{align}
&R_{\text{w}}\geq I(\widetilde{X};A,V)-I(U;Y|A)-I(V;Z|A,U)\label{eq:CSRwOuter}
\end{align}
	such that $U\!-\!V\!-\!(A,\widetilde{X})\!-\!(A,\widetilde{X},X)\!-\!(Y,Z)$ form a Markov chain and $\mathbb{E}[\Gamma(A)]\!\leq\! C$.  It suffices to limit the cardinalities to $|\mathcal{U}|\!\leq\!|\mathcal{A}||\mathcal{\widetilde{X}}|\!+\!3$ and $|\mathcal{V}|\!\leq\!(|\mathcal{A}||\mathcal{\widetilde{X}}|\!+\!3)(|\mathcal{A}||\mathcal{\widetilde{X}}|\!+\!2)$.

\end{theorem}
\begin{IEEEproof}[Proof Sketch]
	The proof for the privacy-leakage rate $R_\ell$ bound uses an inequality that we prove for the CLN channel $(X\geq Z|A,Y)$, which might be useful also for other problems. We also assume another CLN condition $(Z\geq Y|A,X)$ and prove the existence of a single-letter representation of a subtraction of conditional entropies to find a lower bound for the subtraction term by applying the properties of the second CLN condition. The proofs for the secret-key rate, storage rate, and action cost follow by using standard properties of the Shannon entropy.  
\end{IEEEproof}

\begin{remark}
	\normalfont The bounds in Theorem~\ref{theo:ActionBCGSCSOuter} recover the key-leakage-storage-cost regions given for the more restrictive case considered in \cite{bizimKittipongTIFS} with the Markov chain $\widetilde{X}-(A,X)-(Y,Z)$ since we have in (\ref{eq:GSRlOuter}) for the more restrictive case that
	\begin{align}
		&R_\ell\!\geq I(X;A,V,Y) \!-\!I(X;Y|A)\!+\!I(X;Z|A) \nonumber\\
		&\;\qquad+ I(U;Y|A)-I(U;Z|A)\nonumber\\
		&\;\overset{(a)}{=} I(X;A,V,Y) - I(X;Y|A,U) + I(X;Z|A,U)
	\end{align}
	where $(a)$ follows from the Markov chain $U-(A,X)-(Y,Z)$, which is valid only for the more restrictive case. It is straightforward to show that the bounds in Theorem~\ref{theo:ActionBCGSCSOuter} recover the outer bounds given in \cite{IgnaTrans,LaiTrans,bizimITW,bizimWZ}. 
\end{remark}

\section{Example}\label{sec:example}
We illustrate the effects of the storage rate on the achievable secret-key rate and expected action cost. An achievable key-leakage-storage-cost tradeoff with additional assumptions for the auxiliary random variables and with realistic measurement channel parameters suffices to motivate the use of an action in practical biometric secrecy systems and PUFs.
 
Suppose the hidden identifier outputs $X^n$ are uniformly distributed bit sequences, the eavesdropper side information $Z^n$ is a physically-degraded version of the decoder measurements $Y^n$, i.e., $(A,\widetilde{X},X)-Y-Z$ form a Markov chain, and we have binary symmetric channels (BSCs) with crossover probabilities $p$, denoted as BSC$(p)$, such that 
\begin{align}
&P_{\widetilde{X}|X}(\cdot|\cdot)\sim\text{BSC}(p_{\text{enc}})\\ 
&P_{Z|Y}(\cdot|\cdot)\sim\text{BSC}(p_{\text{eve}})\\
&P_{Y|X\widetilde{X}A}(\cdot|\cdot,\widetilde{x},a)\sim\text{BSC}(q_{\widetilde{x}a}) \text{ for all } \widetilde{x},a\in{\{0,1\}}.\label{eq:qxtildea}
\end{align}
Consider physical identifiers like start-up values of static random access memories (SRAM) or RO outputs. Symmetric source outputs and BSCs are realistic source and channel models for such identifiers \cite{SRAMPUFs,bizimMDPI}. We can therefore use the channel parameters obtained in the literature for real ROs and SRAMs.
 
Decoder measurements with smaller crossover probability can be obtained by applying additional post-processing steps, which increases the hardware cost \cite{bizimpaper}. Suppose that the action $A=a$ chooses BSCs such that $q_{\widetilde{x}0}<q_{\widetilde{x}1}$ for all $\widetilde{x}\in\{0,1\}$, i.e., $A=0$ chooses more reliable measurement channels with higher cost. Similarly, assume $q_{0a}<q_{1a}$ for all $a\in\{0,1\}$, i.e., $\widetilde{X}=0$ chooses more reliable measurement channels with higher cost. This assumption is realistic if the ambient temperature increases during encoder measurements since the oscillation frequency of an RO decreases with increasing temperature \cite{bizimtemperaturepaper}, which results in a bias towards the bit $0$ after quantization. In such a case, the decoder measurement channels should be chosen to be more reliable to compensate for the performance loss due to the temperature increase. Moreover, the action cost should be higher for the cases with higher hardware cost. We thus choose the costs of actions as
 \begin{align}
 	&\Gamma(0) = \frac{q_{01}+q_{11}}{q_{01}+q_{11}+q_{10}+q_{00}},\qquad \Gamma(1) = 1-\Gamma(0) \label{eq:choiceofgammasformula}
 \end{align}
where $q_{00}, q_{01}, q_{10}, q_{11}$ are as defined in (\ref{eq:qxtildea}). We use realistic crossover probabilities for RO PUFs combined with the transform-coding algorithm given in \cite{benimdissertation} and satisfy the assumptions given above by choosing $p_{\text{enc}}=0.05$, $q_{00} =0.010$, $q_{10}=0.030$, $q_{01}=0.050$, $q_{11}=0.060$, and $p_{\text{eve}}\approx0.102$ such that $p_{\text{eve}}*q_{11} = 0.150$, where $*$-operator is defined as $p*q=(1-2q)p+q$. The crossover probability of $0.150$ corresponds to the case that the eavesdropper observes noisy PUF ouputs but cannot control the environmental variations as a passive attacker  \cite{bizimtemperaturepaper}. Furthermore, these values result in $\Gamma(0)\approx0.733$ units and $\Gamma(1)\approx0.267$ units by (\ref{eq:choiceofgammasformula}).

\begin{figure}[t]
	\centering
	\newlength\figureheight
	\newlength\figurewidth
	\setlength\figureheight{9.97cm}
	\setlength\figurewidth{4.8cm}
%
%
\begin{tikzpicture}

\begin{axis}[%
width=6.856cm,
height=2.5cm,
at={(0cm,0cm)},
scale only axis,
xmin=0.267,
xmax=0.73,
xlabel style={font=\color{white!15!black}},
xlabel={Cost C},
ymin=0,
ymax=0.32,
ylabel style={font=\color{white!15!black}},
ylabel={$\text{Secret-key Rate R}_\text{s}$},
axis background/.style={fill=white},
xmajorgrids,
ymajorgrids,
legend style={at={(0.49,0.146)}, anchor=south west, legend cell align=left, align=left, draw=white!15!black}
]
\addplot  [color=red, line width=3.0pt]
  table[row sep=crcr]{%
0.267133333333333	0.000754604986853646\\
0.267133333333333	0.000754604986853646\\
0.274133333333333	0.00997817113895916\\
0.290466666666667	0.0305838372071222\\
0.3033	0.0466223249636898\\
0.3278	0.0764529599008656\\
0.358133333333333	0.112404947326869\\
0.3768	0.133911204027946\\
0.3943	0.153517904187038\\
0.4118	0.172451612839989\\
0.431633333333333	0.193475172152709\\
0.449133333333333	0.211118688542058\\
0.465466666666667	0.226995571012191\\
0.473633333333333	0.234541018208105\\
0.486466666666667	0.24622381326917\\
0.5098	0.265752540350576\\
0.528466666666667	0.279479505077079\\
0.535466666666667	0.284016973248991\\
0.556466666666667	0.295268548005922\\
0.561133333333333	0.297701611198381\\
0.5623	0.298103061562843\\
0.563466666666667	0.298489322368883\\
0.564633333333333	0.298859909887597\\
0.5658	0.299214315368461\\
0.566966666666667	0.299552003226017\\
0.568133333333333	0.299872409056417\\
0.5693	0.300174937464074\\
0.570466666666667	0.300458959675952\\
0.571633333333333	0.300723810917827\\
0.5728	0.300968787523134\\
0.573966666666667	0.301193143740662\\
0.575133333333333	0.301396088202254\\
0.5763	0.301576780005644\\
0.577466666666667	0.30173432436047\\
0.578633333333333	0.301867767737067\\
0.5798	0.30197609244759\\
0.580966666666667	0.302058210577066\\
0.582133333333333	0.302112957167544\\
0.585633333333333	0.302112957167544\\
0.5868	0.302112957167544\\
0.592633333333333	0.302112957167544\\
0.727966666666667	0.302112957167544\\
0.729133333333333	0.302112957167544\\
0.7303	0.302112957167544\\
0.731466666666667	0.302112957167544\\
};
\addlegendentry{$\text{R}_\text{w} = 0.001$}

\addplot [color=blue, loosely dashed, line width=3.0pt]
  table[row sep=crcr]{%
0.267133333333333	0.0285441048428285\\
0.267133333333333	0.0285441048428285\\
0.272966666666667	0.0395800749126676\\
0.2753	0.0429031331858865\\
0.2928	0.0643961373937423\\
0.3138	0.0900702903784933\\
0.3348	0.114988423407988\\
0.356966666666667	0.140453125220166\\
0.3803	0.166361693841081\\
0.4013	0.188749417285643\\
0.419966666666667	0.207725229358095\\
0.426966666666667	0.214703319788005\\
0.447966666666667	0.234611008144498\\
0.449133333333333	0.235663935094591\\
0.4678	0.252046058367132\\
0.473633333333333	0.257081978533007\\
0.494633333333333	0.273209460373452\\
0.501633333333333	0.27822062660353\\
0.533133333333333	0.296056920611623\\
0.5343	0.296514848659596\\
0.535466666666667	0.296960899584592\\
0.536633333333333	0.297394764017459\\
0.5378	0.29781611955258\\
0.538966666666667	0.298224629980379\\
0.540133333333333	0.298619944461535\\
0.5413	0.299001696637432\\
0.542466666666667	0.299369503670787\\
0.543633333333333	0.299722965209704\\
0.5448	0.300061662267622\\
0.545966666666667	0.300385156010745\\
0.547133333333333	0.300692986443514\\
0.5483	0.300984670981574\\
0.549466666666667	0.301259702900333\\
0.550633333333333	0.301517549645744\\
0.5518	0.301757650992201\\
0.552966666666667	0.301979417030464\\
0.554133333333333	0.302182225966243\\
0.5553	0.302365421707428\\
0.556466666666667	0.302528311214885\\
0.557633333333333	0.302670161588175\\
0.5588	0.302790196853396\\
0.563466666666667	0.30293822713079\\
0.564633333333333	0.302967962315683\\
0.5658	0.302971231378219\\
0.566966666666667	0.302971231378219\\
0.570466666666667	0.302971231378219\\
0.727966666666667	0.302971231378219\\
0.729133333333333	0.302971231378219\\
0.7303	0.302971231378219\\
0.731466666666667	0.302971231378219\\
};
\addlegendentry{$\text{R}_\text{w} = 0.050$}

\addplot [color=black, densely dotted, line width=3.0pt]
  table[row sep=crcr]{%
0.267133333333333	0.142657119867507\\
0.267133333333333	0.142657119867507\\
0.2683	0.145336712215796\\
0.307966666666667	0.18818445245575\\
0.3103	0.190473970403349\\
0.349966666666667	0.228357374374226\\
0.356966666666667	0.234749055827242\\
0.360466666666667	0.237739294049048\\
0.379133333333333	0.251833309914442\\
0.416466666666667	0.278424405981732\\
0.430466666666667	0.286897377759906\\
0.431633333333333	0.287484670137809\\
0.4328	0.28806572619286\\
0.433966666666667	0.288640450281547\\
0.435133333333333	0.289208744408135\\
0.4363	0.289770508144172\\
0.437466666666667	0.29032563854446\\
0.438633333333333	0.290874030059314\\
0.4398	0.291415574442898\\
0.440966666666667	0.29195016065742\\
0.442133333333333	0.29247767477297\\
0.4433	0.292997999862743\\
0.444466666666667	0.293511015893396\\
0.445633333333333	0.294016599610269\\
0.4468	0.29451462441715\\
0.461966666666667	0.299672412249951\\
0.463133333333333	0.300062561028384\\
0.4643	0.300442959417091\\
0.475966666666667	0.303255713358823\\
0.484133333333333	0.304599279469319\\
0.491133333333333	0.30536054878819\\
0.496966666666667	0.305701581437783\\
0.498133333333333	0.305768633543332\\
0.5028	0.305794666098908\\
0.503966666666667	0.305794666098908\\
0.507466666666667	0.305794666098908\\
0.727966666666667	0.305794666098908\\
0.729133333333333	0.305794666098908\\
0.7303	0.305794666098908\\
0.731466666666667	0.305794666098908\\
};
\addlegendentry{$\text{R}_\text{w} = 0.250$}

\end{axis}
\end{tikzpicture}%
	\caption{Cost vs. secret-key rate projection of the boundary tuples $(R_\text{s},R_\ell,R_{\text{w}},C)$ for the GS model with storage rates $R_\text{w}$ of $0.001, 0.050$, and $0.250$ bits/source-symbol.} \label{fig:costsecretkeycomparisons}
\end{figure}
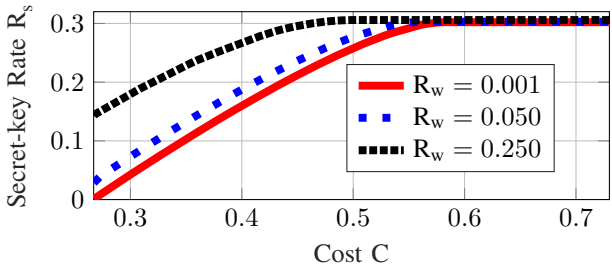

Let the auxiliary random variable $U$ be constant, so from Theorem~\ref{theo:ActionBCgscs} we have an inner bound for the GS model  that is the set of all tuples $(R_\text{s},R_\ell,R_{\text{w}},C)$ satisfying
\begin{align}
&0\leq R_{\text{s}}\leq I(V;Y|A)-I(V;Z|A)\\
&R_\ell \geq\max\Big\{I(V,X;Z|A)+I(X;A,V,Y)-I(V,X;Y|A),\nonumber\\
&\quad\qquad\qquad I(X;A,Z)\Big\}\\
&R_{\text{w}}\geq I(\widetilde{X};A) + I(V;\widetilde{X}|A,Y)
\end{align} 
for some $P_XP_{\widetilde{X}|X}P_{A|\widetilde{X}}P_{YZ|X\widetilde{X}A}P_{V|\widetilde{X}A}$ such that $\mathbb{E}[\Gamma(A)]\!\leq\! C$. Let $P_{A|\widetilde{X}}$ be a binary channel and $P_{V|A\widetilde{X}}(\cdot|a,\cdot)$ be a BSC$(p_a)$ for $a=0,1$. We evaluate an achievable key-leakage-storage-cost region for all possible $P_{A|\widetilde{X}}$ and plot the (cost vs. secret-key rate) projection of the boundary tuples $(R_\text{s},R_\ell,R_{\text{w}},C)$ for different storage rates $R_{\text{w}}$ bits/source-symbol in Fig.~\ref{fig:costsecretkeycomparisons}. Any secret-key rate less than the secret-key rates on the boundary and any cost greater than the cost on the boundary are also achievable. The minimum and maximum expected costs depicted in Fig.~\ref{fig:costsecretkeycomparisons} are $\Gamma(1)$ and $\Gamma(0)$, respectively, since these are costs of two possible actions $A=a$. The secret-key rate $R_\text{s}$ in Fig.~\ref{fig:costsecretkeycomparisons} does not decrease with increasing cost $C$ because for a higher cost better set of channels, i.e., channels with smaller crosssover probabilities, can be used for decoder measurements.

Fig.~\ref{fig:costsecretkeycomparisons} shows that, for a given cost $C=c$, the secret-key rate $R_\text{s}$ increases for increasing storage rate $R_\text{w}$. Moreover, the maximum secret-key rates, denoted by $R_\text{s}^*$, for different storage rates $R_\text{w}$ are different and they are obtained at different minimum cost values, denoted by $C^*$. For instance, for $R_\text{w}\!=\!0.001$ bits/source-symbol we have $(C^*\!=\!0.5821, \;R_\text{s}^*\!=\! 0.3021$ bits/source-symbol); whereas for $R_\text{w}=0.250$ bits/source-symbol we obtain $(C^*\!=\!0.5028, \;R_\text{s}^*\!=\! 0.3058$ bits/source-symbol). This illustrates that having a larger public storage available increases the maximum secret-key rate, e.g., for the given example by approximately $1.22\%$, and significantly decreases the required expected action cost to achieve the maximum secret-key rate, e.g., for this example by approximately $13.62\%$. Thus, a low-complexity PUF design with small hardware area as in \cite{bizimMDPI}, should be used to allocate a large hardware area for helper data storage, which provides significant gains in the achieved rate tuples in combination with an action sequence.

\section{Proof of Theorem~\ref{theo:ActionBCgscs}}\label{sec:InnerBoundsProof}
We provide a proof that follows from the output statistics of random binning (OSRB) method \cite{OSRBAmin} by applying the steps in  \cite[Section 1.6]{BlochLectureNotes2018}. 
\subsection{Proof for the GS Model}

\begin{IEEEproof}[Proof Sketch] Fix $P_{A|\widetilde{X}}$, $P_{V|\widetilde{X}A}$, and $P_{U|V}$ such that $\mathbb{E}[\Gamma(A)]\leq C+\epsilon$ and let $(U^n,V^n,A^n,\widetilde{X}^n,X^n,Y^n,Z^n)$ be i.i.d. according to $P_{UVA\widetilde{X}XYZ}=P_{U|V}P_{V|\widetilde{X}A}P_{A|\widetilde{X}}P_{\widetilde{X}|X}P_XP_{YZ|X\widetilde{X}A}$. Suppose $H(V|U,A,Z)-H(V|U,A,Y)>0$.
	
Assign two random bin indices $(F_{\text{a}},W_{\text{a}})$	to each $a^n$. Assume $F_{\text{a}}\in[1:2^{n\widetilde{R}_{\text{a}}}]$ and $W_{\text{a}}\in[1:2^{nR_{\text{a}}}]$. Similarly, assign two indices $(F_{\text{u}},W_{\text{u}})$ to each $u^n$, where $F_{\text{u}}\in[1:2^{n\widetilde{R}_{\text{u}}}]$ and $W_{\text{u}}\in[1:2^{nR_{\text{u}}}]$. Furthermore, assign three indices $(F_{\text{v}},W_{\text{v}}, S)$ to each $v^n$, where $F_{\text{v}}\in[1:2^{n\widetilde{R}_{\text{v}}}]$, $W_{\text{v}}\in[1:2^{nR_{\text{v}}}]$, and $S\in[1:2^{nR_{\text{s}}}]$. The helper data are $W=(W_{\text{a}}, W_{\text{u}}, W_{\text{v}})$, the public indices are $F=(F_{\text{a}}, F_{\text{u}}, F_{\text{v}})$, and the secret key is $S$.

Reliable estimation of $A^n$ from  $(F_{\text{a}},W_{\text{a}})$ is possible if \cite[Lemma 1]{OSRBAmin}
\begin{align}
	\widetilde{R}_{\text{a}} + R_{\text{a}}> H(A).\label{eq:Anrecons}
\end{align}	
Using a Slepian-Wolf (SW) \cite{SW} decoder, one can reliably estimate $U^n$ from $(F_{\text{u}},W_{\text{u}}, A^n, Y^n)$ if \cite[Lemma 1]{OSRBAmin}
\begin{align}
\widetilde{R}_{\text{u}} + R_{\text{u}}> H(U|Y,A).\label{eq:Unrecons}
\end{align}	
Similarly, 	one can reliably estimate $V^n$ from $(F_{\text{v}},W_{\text{v}}, A^n, Y^n, U^n)$ by using a SW decoder if \cite[Lemma 1]{OSRBAmin}
\begin{align}
\widetilde{R}_{\text{v}} + R_{\text{v}}> H(V|U,Y,A).\label{eq:Vnrecons}
\end{align}
Thus, the reliability constraint in (\ref{eq:reliability_cons}) is satisfied if (\ref{eq:Anrecons})-(\ref{eq:Vnrecons}) are satisfied.
	
The strong secrecy (\ref{eq:secrecyleakage_cons}) and key uniformity (\ref{eq:uniformity_cons}) constraints are satisfied if \cite[Theorem 1]{OSRBAmin}
\begin{align}
	R_\text{s}+\widetilde{R}_{\text{v}} + R_{\text{v}}< H(V|U,A,Z)\label{eq:secrecycond}
\end{align}
since (\ref{eq:secrecycond}) ensures that the three random indices $(S,F_{\text{v}},W_{\text{v}})$ are almost independent of $(U^n,A^n,Z^n)$ and are almost mutually independent and uniformly distributed.
	
The public index $F_{\text{a}}$ is almost independent of $\widetilde{X}^n$, so it is almost independent of $(\widetilde{X}^n,X^n,Y^n,Z^n)$, if we have \cite[Theorem 1]{OSRBAmin}
\begin{align}
	\widetilde{R}_{\text{a}}<H(A|\widetilde{X}).\label{eq:independenceofFa}
\end{align} 
Similarly, the public index $F_{\text{u}}$ is almost independent of $(A^n,\widetilde{X}^n)$, so it is almost independent of $(A^n,\widetilde{X}^n,X^n,Y^n,Z^n)$, if we have \cite[Theorem 1]{OSRBAmin}
\begin{align}
\widetilde{R}_{\text{u}}<H(U|A,\widetilde{X}).\label{eq:independenceofFu}
\end{align} 
Furthermore, the public index $F_{\text{v}}$ is almost independent of $(U^n,A^n,\widetilde{X}^n)$, so it is almost independent of $(U^n,A^n,\widetilde{X}^n,X^n,Y^n,Z^n)$, if we have \cite[Theorem 1]{OSRBAmin}
\begin{align}
\widetilde{R}_{\text{v}}<H(V|U,A,\widetilde{X}).\label{eq:independenceofFv}
\end{align} 
Thus, the public indices $F$ can be fixed by generating them uniformly at random. The encoder can generate $(A^n,U^n,V^n)$ according to $P_{A^nU^nV^n|\widetilde{X}^nF_{\text{a}} F_{\text{u}}F_{\text{v}}}$ obtained from the binning scheme above to compute the bins $W_{\text{a}}$ from $A^n$, $W_{\text{u}}$ from $U^n$, and $(W_{\text{v}},S)$ from $V^n$. This procedure induces a joint probability distribution that is almost equal to $P_{UVA\widetilde{X}XYZ}$ fixed above \cite[Section 1.6]{BlochLectureNotes2018}. 

To satisfy the constraints (\ref{eq:Anrecons})-(\ref{eq:independenceofFv}), we fix the rates to
\begin{align}
&\widetilde{R}_{\text{a}} = H(A|\widetilde{X})-\epsilon\label{eq:R_atildechosen}\\
&R_{\text{a}} = I(A;\widetilde{X})+2\epsilon\label{eq:R_achosen}\\
&\widetilde{R}_{\text{u}} = H(U|A,\widetilde{X})-\epsilon\label{eq:R_utildechosen}\\
&R_{\text{u}} = I(U;\widetilde{X}|A)-I(U;Y|A)+2\epsilon\label{eq:R_uchosen}\\
&\widetilde{R}_{\text{v}} = H(V|U,A,\widetilde{X})-\epsilon\label{eq:R_vtildechosen}\\
&R_{\text{v}} = I(V;\widetilde{X}|A,U)-I(V;Y|A,U)+2\epsilon\label{eq:R_vchosen}\\
&R_{\text{s}} = I(V;Y|A,U)-I(V;Z|A,U)-2\epsilon\label{eq:R_schosen}
\end{align}
for some $\epsilon>0$ such that $\epsilon\rightarrow0$ when $n\rightarrow\infty$. This results in a storage (helper-data) rate $R_{\text{w}}$ of
\begin{align}
&R_{\text{w}} = R_{\text{a}} + R_{\text{u}} + R_{\text{v}}\nonumber\\
	&\;= I(A;\widetilde{X}) + I(U, V;\widetilde{X}|A)-I(U,V;Y|A)+6\epsilon\nonumber\\
	&\; \overset{(a)}{=}  I(A;\widetilde{X}) + I(V;\widetilde{X}|A)-I(V;Y|A)+6\epsilon\nonumber\\
	&\;\overset{(b)}{=}  I(A;\widetilde{X}) + I(V;\widetilde{X}|A,Y)+6\epsilon\label{eq:R_wchosen}
\end{align}
where $(a)$ follows because $U-(V,A)-(\widetilde{X},Y)$ form a Markov chain and $(b)$ follows since $V-(A,\widetilde{X})-Y$ form a Markov chain. Furthermore, since each action sequence $a^n$ is in the typical set with high probability, by the typical average lemma \cite[pp. 26]{Elgamalbook}, the expected cost constraint in (\ref{eq:cost_cons}) is satisfied.

Since $F$ is public, we can bound the privacy leakage as follows.
\begin{align}
	&I(X^n;W,Z^n,F) \overset{(a)}{\leq} I(X^n;W,Z^n|F)+3\epsilon_n\nonumber\\
	&\overset{(b)}{\leq}\!H(X^n|F)\! -\! H(X^n,W,Z^n|F)\!+\!H(W|F)\!+\!H(Z^n|A^n,U^n)\nonumber\\
	&\quad \!+\!I(A^n,U^n;Z^n|W_{\text{a}},F_{\text{a}},W_{\text{u}},F_{\text{u}})\!+\!4\epsilon_n\nonumber\\
	&\overset{(c)}{\leq} \!-H(Z^n|X^n,F) \!-\! H(W,A^n|X^n,Z^n,F)+H(W|F)\nonumber\\
	&\quad\!+\!H(A^n|W,X^n,Z^n,F)+H(Z^n|A^n,U^n)\!+\! 4\epsilon_n\!+\!n\epsilon_n^{\prime}\nonumber\\
	&\quad + I(U^n;Z^n|A^n,W_\text{u},F_\text{u})\nonumber\\
	&\overset{(d)}{\leq}\!-\!H(A^n|X^n,F_\text{a})\!+\!I(F_\text{u},F_\text{v};A^n|X^n,F_\text{a})\! -\! H(Z^n|A^n,X^n)\nonumber\\
	&\quad\! +\!I(Z^n;F_\text{u},F_\text{v}|A^n,X^n)\!-\! H(W|A^n,X^n,Z^n,F)\!+\!2n\epsilon_n^{\prime}\nonumber\\
	&\quad \!+\! 4\epsilon_n \!+\! H(W|F)\!+\!H(Z^n|A^n,U^n)\!+\! I(U^n;Z^n|A^n,W_\text{u},F_\text{u})\nonumber\\
	&\overset{(e)}{\leq} \!-\!H(A^n|X^n)\!+\!H(F_\text{a}|X^n)\!-\!H(Z^n|A^n,X^n)\!+\!2n\epsilon_n^{\prime}\!+\!8\epsilon_n\nonumber\\
	&\quad -H(W_\text{u},V^n|A^n,X^n,Z^n,F)+H(V^n|W,A^n,X^n,Z^n,F)\nonumber\\
	&\quad  \!+\! H(W|F)\!+\!H(Z^n|A^n,U^n)\!+\! I(U^n;Z^n|A^n,W_\text{u},F_\text{u})\nonumber\\
	&\overset{(f)}{\leq} \!-\!H(A^n|X^n)\!+\!nH(A|\widetilde{X})\!-\!H(Z^n|A^n,X^n)\!+\!2n\epsilon_n^{\prime}\!+\!8\epsilon_n\nonumber\\
	&\quad \!-\!H(V^n|A^n,\widetilde{X}^n,F)\!-\!I(V^n;\widetilde{X}^n|A^n,X^n,Z^n,F)\nonumber\\
	&\quad\!+\!H(V^n|W,A^n,X^n,Z^n,F)\!+\! H(W|F)\!+\!H(Z^n|A^n,U^n)\nonumber\\
	&\quad \!+\!nI(U;Y|A)+5\epsilon_n\!-\!H(U^n|Z^n,A^n,W_\text{u},F_\text{u})\nonumber\\
	&\overset{(g)}{\leq}\!-\!H(A^n|X^n)\!+\!nH(A|\widetilde{X})\!-\!H(Z^n|A^n,X^n,U^n)\nonumber\\
	&\quad\!-\!I(U^n;Z^n|A^n,X^n)\!+\!2n\epsilon_n^{\prime}\!+\!15\epsilon_n\nonumber\\
	&\quad \!-\!H(\widetilde{X}^n|A^n,X^n,Z^n)\!+\!H(\widetilde{X}^n|A^n,X^n,Z^n,V^n)\nonumber\\
	&\quad \!+\!H(V^n|W,A^n,X^n,Z^n,F)\!+\! H(W_\text{a})\!+\! H(W_\text{u})\!+\! H(W_\text{v})\nonumber\\
	&\quad\!+\!H(Z^n|A^n,U^n)\!+\!nI(U;Y|A)\!-\!H(U^n|Z^n,A^n,W_\text{u},F_\text{u})\nonumber\\
	&\overset{(h)}{\leq}\!n\big(\!-\!H(A|X)\!+\!H(A|\widetilde{X})\!+\! I(X;Z|A,U)\!-\!I(U;Z|A,X)\big)\nonumber\\
	&\quad -nI(\widetilde{X};V|A,X,Z)\!+\!2n\epsilon_n^{\prime}\!+\!15\epsilon_n\nonumber\\
	&\quad+n\big(I(\widetilde{X};A,V)-I(U;Y|A)-I(V;Y|A,U)+6\epsilon\big)\nonumber\\
	&\quad +nI(U;Y|A)-H(U^n|Z^n,A^n,W_\text{u},F_\text{u})\nonumber\\
	&\quad +\!H(V^n|W,A^n,X^n,Z^n,F)\nonumber\\
	&\overset{(i)}{\leq}\!n\big(I(X;Z|A,U)\!-\!I(U;Z|A,X)\big)\!+\!2n\epsilon_n^{\prime}\!+\!15\epsilon_n+6n\epsilon \nonumber\\
	&\quad \!+n\Big(\!-\!H(V,A|X)\!+\!I(V;Z|A,X)\!+\!H(V,A|\widetilde{X})\!\Big)\nonumber\\
	&\quad+n\big(I(\widetilde{X};A,V)-I(V;Y|A,U)\big)\nonumber\\
	&\quad \!-\!H(U^n|Z^n,A^n,W_\text{u},F_\text{u}) +\!H(V^n|W,A^n,X^n,Z^n,F)\nonumber\\
	&\overset{(j)}{=}n\big(I(X;Z|A,U)+I(X;A,V)-I(V;Y|A,U)  \big)\nonumber\\
	&\quad \!+\!2n\epsilon_n^{\prime}\!+\!15\epsilon_n+6n\epsilon+nI(V;Z|A,X,U)\nonumber\\
	&\quad \!-\!H(U^n|Z^n,A^n,W_\text{u},F_\text{u}) \!+\!H(V^n|W,A^n,X^n,Z^n,F)\nonumber
					\end{align}
	\begin{align}
	&\overset{(k)}{=}n\big(I(X;Z|A,U)+I(X;A,V,Y)-I(X,V;Y|A,U)  \big) \nonumber\\
	&\quad \!+\!2n\epsilon_n^{\prime}\!+\!15\epsilon_n+6n\epsilon+nI(V;Z|A,X,U)\nonumber\\
	&\quad \!-\!H(U^n|Z^n,A^n,W_\text{u},F_\text{u})\!+\!H(V^n|W,A^n,X^n,Z^n,F)\nonumber\\
	&=n\big(I(X;Z|A,U)+I(X;A,V,Y)-I(X;Y|A,U)  \big) \nonumber\\
	&\quad \!+\!2n\epsilon_n^{\prime}\!+\!15\epsilon_n+6n\epsilon+nI(V;Z|A,X,U)\nonumber\\
	&\quad \!-\!H(U^n|Z^n,A^n,W_\text{u},F_\text{u})-nI(V;Y|A,X,U)\nonumber\\
	&\quad \!+\!H(V^n|W,A^n,X^n,Z^n,F)\label{eq:achleakfirstpart}
\end{align} 
where \\
$(a)$ follows from \cite[Theorem 1]{OSRBAmin} such that we have that $F$ is almost independent of $(\widetilde{X}^n,X^n)$ due to the Markov chain $X^n - \widetilde{X}^n - F$. We have 
\begin{align}
	&I(X^n;F) = I(X^n;F_\text{a})+I(X^n;F_\text{u}|F_\text{a})+I(X^n;F_\text{v}|F_\text{a},F_\text{u})\nonumber\\
	&\;\overset{(a.1)}{\leq} 3\epsilon_n\label{eq:XnandFareind}
\end{align}
where $(a.1)$ follows by (\ref{eq:R_atildechosen}), (\ref{eq:R_utildechosen}), and (\ref{eq:R_vtildechosen}), and from the facts that $A^n$ determines $F_\text{a}$ and $U^n$ determines $F_\text{u}$, for some $\epsilon_n>0$ with $\epsilon_n\rightarrow 0$ when $n\rightarrow\infty$;\\
$(b)$ follows because $(F_\text{v},W_\text{v})$ are almost independent of $(U^n,A^n,Z^n)$ by (\ref{eq:secrecycond}), $A^n$ determines $(W_\text{a},F_\text{a})$, and $U^n$ determines $(W_\text{u},F_\text{u})$;\\
$(c)$ follows since $A^n$ determines $(F_\text{a},W_\text{a})$, and from (\ref{eq:Anrecons}) and \cite[Lemma 1]{OSRBAmin} we have $H(A^n|W_\text{a},F_\text{a})\leq n\epsilon_n^{\prime}$ for some $\epsilon_n^{\prime}\rightarrow 0$ when $n\rightarrow\infty$;\\
$(d)$ follows because by (\ref{eq:Anrecons}) such that $H(A^n|W,X^n,Z^n,F)\leq n\epsilon_n^{\prime}$ for some $\epsilon_n^{\prime}\rightarrow 0$ when $n\rightarrow\infty$ and $A^n$ determines $F_\text{a}$;\\
$(e)$ follows by (\ref{eq:independenceofFu}) and (\ref{eq:independenceofFv}), and because $A^n$ determines $W_\text{a}$, $V^n$ determines $W_\text{v}$, and we obtain
\begin{align*}
	&I(F_\text{u},F_\text{v};A^n|X^n,F_\text{a})\nonumber\\
	& \;= I(F_\text{u};A^n|X^n,F_\text{a}) +I(F_\text{v};A^n|X^n,F_\text{a},F_\text{u})\leq 2\epsilon_n
\end{align*}
and
\begin{align*}
&I(Z^n;F_\text{u},F_\text{v}|A^n,X^n)\nonumber\\
& \;= I(Z^n;F_\text{u}|A^n,X^n) +I(Z^n;F_\text{v}|A^n,X^n,F_\text{u})\leq 2\epsilon_n
\end{align*}
since $F_\text{u}$ is almost independent of $(A^n,X^n,Z^n)$, $U^n$ determines $F_\text{u}$, and  $F_\text{v}$ is almost independent of $(U^n,A^n,X^n,Z^n)$;\\
$(f)$ follows by (\ref{eq:R_atildechosen}), from the Markov chain $V^n-(A^n,\widetilde{X}^n,F)-(X^n,Z^n)$, and from the following inequality
\begin{align}
	&H(U^n|A^n,W_\text{u},F_\text{u}) \overset{(f.1)}{=} H(U^n|A^n)-H(W_\text{u},F_\text{u}|A^n)\nonumber\\
    &\;\overset{(f.2)}{\leq} H(U^n|A^n)-(H(W_\text{u}|A^n)+H(F_\text{u}|A^n)-\epsilon_n)\nonumber\\
    &\;\overset{(f.3)}{\leq} H(U^n|A^n)-(H(W_\text{u})-\epsilon_n)-(H(F_\text{u})-\epsilon_n)+\epsilon_n\nonumber\\
    &\;\overset{(f.4)}{\leq} H(U^n|A^n)-(nR_\text{u}-\epsilon_n)-(n\widetilde{R}_\text{u}-\epsilon_n)+3\epsilon_n\nonumber\\
    &\;\overset{(f.5)}{\leq} nI(U;Y|A)+5\epsilon_n
\end{align}
where $(f.1)$ follows because $U^n$ determines $(W_\text{u},F_\text{u})$, $(f.2)$ follows from $R_\text{u}+\widetilde{R}_\text{u}<H(U|A)$ for some $\epsilon_n>0$ such that $\epsilon_n\rightarrow0$ when $n\rightarrow\infty$, $(f.3)$ follows because $W_\text{u}$ is almost independent of $A^n$ due to $R_\text{u}<H(U|A)$ when $n\rightarrow\infty$ and $F_\text{u}$ is almost independent of $A^n$ due to $\widetilde{R}_\text{u}<H(U|A)$ when $n\rightarrow\infty$ for some $\epsilon_n>0$ such that $\epsilon_n\rightarrow0$ when $n\rightarrow\infty$, $(f.4)$ follows because $(W_\text{u},F_\text{u})$ are almost uniformly distributed due to $R_\text{u}+\widetilde{R}_\text{u}<H(U)$, and $(f.5)$ follows by (\ref{eq:R_utildechosen}) and (\ref{eq:R_uchosen});\\
$(g)$ follows since $A^n$ determines $F_\text{a}$ and by (\ref{eq:independenceofFu}) and (\ref{eq:independenceofFv}), and we obtain
\begin{align*}
	&I(F_\text{u},F_\text{v};\widetilde{X}^n|A^n,X^n,Z^n)\nonumber\\
	&\; \!=\! I(F_\text{u};\widetilde{X}^n|A^n,X^n,Z^n)\!+\!I(F_\text{v};\widetilde{X}^n|A^n,X^n,Z^n,F_\text{u})\leq 2\epsilon_n
\end{align*}
since $F_\text{u}$ is almost independent of $(A^n,X^n,Z^n,\widetilde{X}^n)$, $U^n$ determines $F_\text{u}$, and  $F_\text{v}$ is almost independent of $(U^n,A^n,X^n,Z^n,\widetilde{X}^n)$;\\
$(h)$ follows by (\ref{eq:R_achosen}), (\ref{eq:R_uchosen}), and (\ref{eq:R_vchosen}), and from the Markov chains $A-\widetilde{X}-X$ and $U-(V,A)-\widetilde{X}$;\\
$(i)$ follows from the Markov chain $V-(A,\widetilde{X})-(X,Z)$;\\
$(j)$ follows from the Markov chain $Z-(A,X,V)-U$;\\
$(k)$ follows from the Markov chain $U-(V,A)-(Y,X)$. 

Consider multi-letter terms in (\ref{eq:achleakfirstpart}), i.e., $-H(U^n|Z^n,A^n,W_\text{u},F_\text{u})\!+\!H(V^n|W,A^n,X^n,Z^n,F)$. There are nine cases we have to analyze for this sum.

\textbf{Case 1}: Suppose we have
\begin{align}
&R_\text{u}+\widetilde{R}_\text{u} < H(U|Z,A,X) \label{eq:case1_1}\\
&R_\text{v}+\widetilde{R}_\text{v} \geq H(V|Z,A,X)\label{eq:case1_2}.
\end{align}
Then, $W_\text{u}$, $F_\text{u}$, and $(Z^n,A^n,X^n)$ are almost mutually independent, and $W_\text{u}$ and $F_\text{u}$ are almost uniformly distributed by \cite[Theorem 1]{OSRBAmin}. Furthermore, we can recover $V^n$ from $(F_\text{v},W_\text{v},Z^n,A^n,X^n)$ by using a SW decoder \cite[Lemma 1]{OSRBAmin}. We obtain
\begin{align}
	&-H(U^n|Z^n,A^n,W_\text{u},F_\text{u})\!+\!H(V^n|W,A^n,X^n,Z^n,F)\nonumber\\
	&\; \overset{(a)}{\leq} -H(U^n|Z^n,A^n) + H(W_\text{u},F_\text{u}|Z^n,A^n)\nonumber\\
	&\;\qquad +H(V^n|W_\text{v},A^n,X^n,Z^n,F_\text{v})\nonumber\\
	&\; \overset{(b)}{\leq}-H(U^n|Z^n,A^n) + H(W_\text{u}) +H(F_\text{u})+n\epsilon_n^{\prime}\nonumber\\
	&\;   \overset{(c)}{\leq}n(-H(U|Z,A) + H(U|Y,A)+\epsilon+\epsilon_n^{\prime})\label{eq:case1step1}
\end{align}
where $(a)$ follows because $U^n$ determines $(F_\text{u},W_\text{u})$, $(b)$ follows by (\ref{eq:case1_2}) such that $H(V^n|F_\text{v},W_\text{v},Z^n,A^n,X^n)\leq n\epsilon^{\prime}_n$ for some $\epsilon_n^{\prime}>0$ such that $\epsilon_n^{\prime}\rightarrow 0$ when $n\rightarrow\infty$, and $(c)$ follows by (\ref{eq:R_utildechosen}) and (\ref{eq:R_uchosen}).

Combining (\ref{eq:achleakfirstpart}) and (\ref{eq:case1step1}), we obtain for Case 1
\begin{align}
&I(X^n;W,Z^n,F)\nonumber\\
&\; \overset{(a)}{\leq}  n\big(I(X;Z|A,U)+I(X;A,V,Y)-I(X;Y|A,U)  \big)\nonumber\\
&\quad +n\big(I(V;Z|A,X,U)\!-\!I(V;Y|A,X,U)\big)\nonumber\\
&\quad +n(I(U;Z|A) -I(U;Y|A)+\epsilon^{\prime\prime})\nonumber\\
&\;\overset{(b)}{=} n\big(I(V,X;Z|A)+I(X;A,V,Y)-I(V,X;Y|A)  \big)\nonumber\\
&\quad +n\epsilon^{\prime\prime}\label{eq:case1privleak}
\end{align}
where $(a)$ follows for some $\epsilon^{\prime\prime}>0$ such that $\epsilon^{\prime\prime}\rightarrow 0$ when $n\rightarrow\infty$ and $(b)$ follows from the Markov chain $U-V-(A,X,Z,Y)$.

\textbf{Case 2}: Suppose we have
\begin{align}
&R_\text{u}+\widetilde{R}_\text{u} < H(U|Z,A,X) \label{eq:case2_1}\\
&R_\text{v}+\widetilde{R}_\text{v} < H(V|Z,A,X)\label{eq:case2_2}\\
&R_\text{v}+\widetilde{R}_\text{v} \geq H(V|U,Z,A,X)\label{eq:case2_3}.
\end{align}
Then, $W_\text{u}$, $F_\text{u}$, and $(Z^n,A^n,X^n)$ are almost mutually independent, and $W_\text{u}$ and $F_\text{u}$ are almost uniformly distributed by \cite[Theorem 1]{OSRBAmin}. Similarly, $W_\text{v}$, $F_\text{v}$, and $(Z^n,A^n,X^n)$ are almost mutually independent  \cite[Theorem 1]{OSRBAmin}, but we can recover $V^n$ from $(U^n,Z^n,A^n,X^n)$ \cite[Lemma 1]{OSRBAmin}. Moreover, $W_\text{v}$ and $F_\text{v}$ are almost uniformly distributed \cite[Theorem 1]{OSRBAmin}. We have
\begin{align}
&-H(U^n|Z^n,A^n,W_\text{u},F_\text{u})\!+\!H(V^n|W,A^n,X^n,Z^n,F)\nonumber\\
&\; \overset{(a)}{\leq} -nH(U|Z,A) + H(W_\text{u},F_\text{u}|Z^n,A^n)+nH(V|A,X,Z)\nonumber\\
&\;\quad -H(W_\text{u} ,F_\text{u}|A^n,X^n,Z^n) \nonumber\\
&\;\quad -H(W_\text{v},F_\text{v}|A^n,X^n,Z^n,W_\text{u} ,F_\text{u}) +nH(U|A,X,Z,V)\nonumber\\
&\; \overset{(b)}{\leq} - nH(U|Z,A) \!+\! H(W_\text{u})+H(F_\text{u})\!+\!nH(U,V|A,X,Z)\nonumber\\
&\;\quad -(H(W_\text{u}|A^n,X^n,Z^n) +H(F_\text{u}|A^n,X^n,Z^n)-\epsilon_n) \nonumber\\
&\;\quad -H(F_\text{v}|A^n,X^n,Z^n,W_\text{u} ,F_\text{u}) \nonumber\\
&\;\quad -H(W_\text{v}|A^n,X^n,Z^n,W_\text{u} ,F_\text{u},F_\text{v}) \nonumber\\
&\; \overset{(c)}{\leq}  -nH(U|Z,A)\!+\! H(W_\text{u})+H(F_\text{u})\!+\!nH(U,V|A,X,Z)\nonumber\\
&\;\quad -(H(W_\text{u})\!-\!\epsilon_n) -(H(F_\text{u})\!-\!\epsilon_n)+\epsilon_n \nonumber\\
&\;\quad -(H(F_\text{v})-\epsilon_n)-H(W_\text{v}|A^n,X^n,Z^n,U^n,F_\text{v}) \nonumber\\
&\; \overset{(d)}{\leq} -nH(U|Z,A) +\!nH(U,V|A,X,Z)\nonumber\\
&\;\quad +4\epsilon_n  -H(F_\text{v})-H(V^n|A^n,X^n,Z^n,U^n,F_\text{v})+n\epsilon^{\prime}_n\nonumber\\
&\; \overset{(e)}{\leq} -nH(U|Z,A) +\!nH(U,V|A,X,Z)\nonumber\\
&\;\quad +4\epsilon_n  -H(V^n|A^n,X^n,Z^n,U^n)\!+\!n\epsilon^{\prime}_n\nonumber\\
&\; = -nI(U;X|Z,A)+4\epsilon_n+n\epsilon^{\prime}_n
\label{eq:case2step1}
\end{align}
where $(a)$ follows because $U^n$ determines $(F_\text{u},W_\text{u})$, $A^n$ determines $(F_\text{a},W_\text{a})$, and $V^n$ determines $(F_\text{v},W_\text{v})$, $(b)$ follows by (\ref{eq:case2_1}) such that $W_\text{u}$ and $F_\text{u}$ are almost independent given $(A^n,X^n,Z^n)$, $(c)$ follows by (\ref{eq:case2_1}) such that $W_\text{u}$ and $F_\text{u}$ are almost independent of $(A^n,X^n,Z^n)$ and by (\ref{eq:independenceofFv}) such that $F_\text{v}$ is almost independent of $(A^n,X^n,Z^n,U^n)$ and $U^n$ determines $W_\text{u}$ and $F_\text{u}$, $(d)$ follows because $V^n$ determines $W_\text{v}$ and by (\ref{eq:case2_3}), and $(e)$ follows since $V^n$ determines $F_\text{v}$.

Combining (\ref{eq:achleakfirstpart}) and (\ref{eq:case2step1}), we obtain for Case 2
\begin{align} 
&I(X^n;W,Z^n,F)\nonumber\\
&\; \overset{(a)}{\leq}  n\big(I(X;Z|A,U)+I(X;A,V,Y)-I(X;Y|A,U)  \big)\nonumber\\
&\quad +n\big(I(V;Z|A,X,U)\!-\!I(V;Y|A,X,U)\big)\nonumber\\
&\quad -n(I(U;X|Z,A)+\!\epsilon^{\prime\prime})\nonumber\\
&\; = n\big(I(V,X;Z|A,U)+I(X;A,V,Y)-I(V,X;Y|A,U)  \big)\nonumber\\
&\quad -n(I(U;X|Z,A)+\!\epsilon^{\prime\prime})\label{eq:case2privleak}
\end{align}
where $(a)$ follows for some $\epsilon^{\prime\prime}>0$ such that $\epsilon^{\prime\prime}\rightarrow 0$ when $n\rightarrow\infty$.

\textbf{Case 3}: Suppose we have
\begin{align}
&R_\text{u}+\widetilde{R}_\text{u} < H(U|Z,A,X) \label{eq:case3_1}\\
&R_\text{v}+\widetilde{R}_\text{v} < H(V|U,Z,A,X)\label{eq:case3_2}.
\end{align}
Then, $W_\text{u}$, $F_\text{u}$, and $(Z^n,A^n,X^n)$ are almost mutually independent, and $W_\text{u}$ and $F_\text{u}$ are almost uniformly distributed by \cite[Theorem 1]{OSRBAmin}. Similarly, $W_\text{v}$, $F_\text{v}$, and $(U,^nZ^n,A^n,X^n)$ are almost mutually independent, and $W_\text{v}$ and $F_\text{v}$ are almost uniformly distributed by \cite[Theorem 1]{OSRBAmin}. We have
\begin{align}
&-H(U^n|Z^n,A^n,W_\text{u},F_\text{u})\!+\!H(V^n|W,A^n,X^n,Z^n,F)\nonumber\\
&\; \overset{(a)}{\leq} -nH(U|Z,A) + H(W_\text{u},F_\text{u}|Z^n,A^n)+nH(V|A,X,Z)\nonumber\\
&\;\quad -H(W_\text{u} ,F_\text{u}|A^n,X^n,Z^n) \nonumber\\
&\;\quad -H(W_\text{v},F_\text{v}|A^n,X^n,Z^n,W_\text{u} ,F_\text{u}) \nonumber\\
&\;\quad +nH(U|A,X,Z,V)\nonumber\\
&\; \overset{(b)}{\leq} - nH(U|Z,A) \!+\! H(W_\text{u})+H(F_\text{u})\!+\!nH(U,V|A,X,Z)\nonumber\\
&\;\quad -(H(W_\text{u}|A^n,X^n,Z^n) +H(F_\text{u}|A^n,X^n,Z^n)-\epsilon_n) \nonumber\\
&\;\quad -H(F_\text{v}|A^n,X^n,Z^n,W_\text{u} ,F_\text{u}) \nonumber\\
&\;\quad -H(W_\text{v}|A^n,X^n,Z^n,W_\text{u} ,F_\text{u},F_\text{v}) \nonumber\\
&\; \overset{(c)}{\leq}  -nH(U|Z,A)\!+\! H(W_\text{u})+H(F_\text{u})\!+\!nH(U,V|A,X,Z)\nonumber\\
&\;\quad -(H(W_\text{u})\!-\!\epsilon_n) -(H(F_\text{u})\!-\!\epsilon_n)+\epsilon_n \nonumber\\
&\;\quad -(H(F_\text{v})-\epsilon_n)-H(W_\text{v}|A^n,X^n,Z^n,U^n,F_\text{v}) \nonumber\\
&\; = -nH(U|Z,A) +\!nH(U,V|A,X,Z)\nonumber\\
&\;\quad +4\epsilon_n  -H(F_\text{v})-H(V^n|A^n,X^n,Z^n,U^n,F_\text{v})\nonumber\\
&\;\quad +H(V^n|A^n,X^n,Z^n,U^n)\nonumber\\
&\;\quad-I(V^n;W_\text{v},F_\text{v}|A^n,X^n,Z^n,U^n)\nonumber\\
&\; \overset{(d)}{\leq} -nH(U|Z,A) +\!nH(U,V|A,X,Z)\nonumber\\
&\;\quad +4\epsilon_n  -H(F_\text{v})-H(V^n|A^n,X^n,Z^n,U^n)+H(F_\text{v})\nonumber\\
&\;\quad +H(V^n|A^n,X^n,Z^n,U^n)\nonumber\\
&\;\quad\!-\!(H(W_\text{v}|A^n,X^n,Z^n,U^n)\!+\!H(F_\text{v}|A^n,X^n,Z^n,U^n)\!-\!\epsilon_n)\nonumber\\
&\; \overset{(e)}{\leq} -nH(U|Z,A) +\!nH(U,V|A,X,Z)\nonumber\\
&\;\quad +4\epsilon_n\!-\!(nR_\text{v}-2\epsilon_n)\!-(n\widetilde{R}_\text{v}-2\epsilon_n)\!+\!\epsilon_n\nonumber\\
&\; \overset{(f)}{\leq} -nI(U;X|Z,A) +9\epsilon_n\nonumber\\
&\;\quad+ n(I(V;Y|A,U)-I(V;X,Z|A,U))\label{eq:case3step1}
\end{align}
where $(a)$ follows because $U^n$ determines $(F_\text{u},W_\text{u})$, $A^n$ determines $(F_\text{a},W_\text{a})$, and $V^n$ determines $(F_\text{v},W_\text{v})$, $(b)$ follows by (\ref{eq:case2_1}) such that $W_\text{u}$ and $F_\text{u}$ are almost independent given $(A^n,X^n,Z^n)$, $(c)$ follows by (\ref{eq:case2_1}) such that $W_\text{u}$ and $F_\text{u}$ are almost independent of $(A^n,X^n,Z^n)$ and by (\ref{eq:independenceofFv}) such that $F_\text{v}$ is almost independent of $(A^n,X^n,Z^n,U^n)$ and $U^n$ determines $W_\text{u}$ and $F_\text{u}$, $(d)$ follows by (\ref{eq:case3_2}) and since $V^n$ determines $(W_\text{v},F_\text{v})$, $(e)$ follows because $W_\text{v}$ and $F_\text{v}$ are almost independent of $(A^n,X^n,Z^n,U^n)$ and are almost uniformly distributed, and $(f)$ follows by (\ref{eq:R_vtildechosen}) and (\ref{eq:R_vchosen}).

Combining (\ref{eq:achleakfirstpart}) and (\ref{eq:case3step1}), we obtain for Case 3
\begin{align} 
&I(X^n;W,Z^n,F)\nonumber\\
&\; \overset{(a)}{\leq}  n\big(I(X;Z|A,U)+I(X;A,V,Y)-I(X;Y|A,U)  \big)\nonumber\\
&\;\quad +n\big(I(V;Z|A,X,U)\!-\!I(V;Y|A,X,U)+\!\epsilon^{\prime\prime})\big)\nonumber\\
&\;\quad +n(-I(U;X|Z,A) + I(V;Y|A,U)-I(V;X,Z|A,U))\nonumber\\
&\; = n\big(I(V,X;Z|A,U)+I(X;A,V,Y)-I(V,X;Y|A,U)  \big)\nonumber\\
&\;\quad \!+\!n(\!-\!I(U;X|Z,A) \!+\! I(V;Y|A,U)\!-\!I(V;X,Z|A,U)\!+\!\epsilon^{\prime\prime})\nonumber\\
&\; \overset{(b)}{=} n(I(V;Z|A,U)+I(X;Z|A,V)+I(X;A,V,Y))\nonumber\\
&\;\quad +n(-I(X;Y|A,V)-I(U;X|Z,A)-I(V;Z|A,U))\nonumber\\
&\;\quad -nI(V;X|A,U,Z) + n\epsilon^{\prime\prime}\nonumber\\
&\;\overset{(c)}{=} nI(X;A,V,Z)-nI(V;X|Z,A)+ n\epsilon^{\prime\prime}\nonumber\\
&\;= nI(X;A,Z)+n\epsilon^{\prime\prime}\label{eq:case3privleak}
\end{align}
where $(a)$ follows for some $\epsilon^{\prime\prime}>0$ such that $\epsilon^{\prime\prime}\rightarrow 0$ when $n\rightarrow\infty$, $(b)$ follows from the Markov chain $U-V-(A,X,Y,Z)$, and $(c)$ follows from the Markov chain $U-V-(X,A,Z)$.

\textbf{Case 4}: Suppose we have
\begin{align}
&R_\text{u}+\widetilde{R}_\text{u} \geq H(U|Z,A,X)\label{eq:case4_1}\\
&R_\text{u}+\widetilde{R}_\text{u} < H(U|Z,A) \label{eq:case4_2}\\
&R_\text{v}+\widetilde{R}_\text{v} \geq H(V|U,Z,A,X)\label{eq:case4_3}\\
&R_\text{v}+\widetilde{R}_\text{v} < H(V|Z,A,X)\label{eq:case4_4}.
\end{align}
Then, $W_\text{u}$, $F_\text{u}$, and $(Z^n,A^n)$ are almost mutually independent, and $W_\text{u}$ and $F_\text{u}$ are almost uniformly distributed by \cite[Theorem 1]{OSRBAmin} and we can recover $U^n$ from $(F_\text{u},W_\text{u},Z^n,A^n,X^n)$ by using a SW decoder  \cite[Lemma 1]{OSRBAmin}. Moreover, we can recover $V^n$ from $(F_\text{v},W_\text{v},U^n, Z^n,A^n,X^n)$ by using a SW decoder \cite[Lemma 1]{OSRBAmin}, but $F_\text{v}$, $W_\text{v}$, and $(Z^n,A^n,X^n)$ are almost independent, and $F_\text{v}$ and $W_\text{v}$ are almost uniformly distributed by \cite[Theorem 1]{OSRBAmin}. We obtain
\begin{align}
&-H(U^n|Z^n,A^n,W_\text{u},F_\text{u})\!+\!H(V^n|W,A^n,X^n,Z^n,F)\nonumber\\
&\; \overset{(a)}{\leq} -H(U^n|Z^n,A^n) + H(W_\text{u},F_\text{u}|Z^n,A^n)\nonumber\\
&\;\qquad +H(V^n|W_\text{v},U^n,A^n,X^n,Z^n,F_\text{v})\nonumber\\
&\;\qquad+H(U^n|W_\text{u},F_\text{u},A^n,X^n,Z^n)\nonumber\\
&\; \overset{(b)}{\leq}-H(U^n|Z^n,A^n) + H(W_\text{u}) +H(F_\text{u})+2n\epsilon_n^{\prime}\nonumber\\
&\;   \overset{(c)}{\leq}n(-H(U|Z,A) + H(U|Y,A)+\epsilon+2\epsilon_n^{\prime})\label{eq:case4step1}
\end{align}
where $(a)$ follows because $U^n$ determines $(F_\text{u},W_\text{u})$ and $A^n$ determines $(F_\text{a},W_\text{a})$, $(b)$ follows by (\ref{eq:case4_1}) and (\ref{eq:case4_3}) for some $\epsilon_n^{\prime}>0$ such that $\epsilon_n^{\prime}\rightarrow 0$ when $n\rightarrow\infty$, and $(c)$ follows by (\ref{eq:R_utildechosen}) and (\ref{eq:R_uchosen}).

Combining (\ref{eq:achleakfirstpart}) and (\ref{eq:case4step1}), we obtain for Case 4
\begin{align}
&I(X^n;W,Z^n,F)\nonumber\\
&\; \overset{(a)}{\leq}  n\big(I(X;Z|A,U)+I(X;A,V,Y)-I(X;Y|A,U)  \big)\nonumber\\
&\quad +n\big(I(V;Z|A,X,U)\!-\!I(V;Y|A,X,U)\big)\nonumber\\
&\quad +n(I(U;Z|A) -I(U;Y|A)+\epsilon^{\prime\prime})\nonumber\\
&\; \overset{(b)}{=} n\big(I(V,X;Z|A)+I(X;A,V,Y)-I(V,X;Y|A)  \big)\nonumber\\
&\quad +n\epsilon^{\prime\prime}\label{eq:case4privleak}
\end{align}
where $(a)$ is for some $\epsilon^{\prime\prime}>0$ such that $\epsilon^{\prime\prime}\rightarrow 0$ when $n\rightarrow\infty$ and $(b)$ follows from the Markov chain $U-V-(A,X,Z,Y)$.

\textbf{Case 5}: Suppose we have
\begin{align}
&R_\text{u}+\widetilde{R}_\text{u} \geq H(U|Z,A,X)\label{eq:case5_1}\\
&R_\text{u}+\widetilde{R}_\text{u} < H(U|Z,A) \label{eq:case5_2}\\
&R_\text{v}+\widetilde{R}_\text{v} < H(V|U,Z,A,X)\label{eq:case5_3}.
\end{align}
Then, $W_\text{u}$, $F_\text{u}$, and $(Z^n,A^n)$ are almost mutually independent, and $W_\text{u}$ and $F_\text{u}$ are almost uniformly distributed by \cite[Theorem 1]{OSRBAmin} and we can recover $U^n$ from $(F_\text{u},W_\text{u},Z^n,A^n,X^n)$ by using a SW decoder  \cite[Lemma 1]{OSRBAmin}. Moreover, $W_\text{v}$, $F_\text{v}$, and $(U^n,Z^n,A^n,X^n)$ are almost mutually independent, and $W_\text{v}$ and $F_\text{v}$ are almost uniformly distributed \cite[Theorem 1]{OSRBAmin}. We have
\begin{align}
&-H(U^n|Z^n,A^n,W_\text{u},F_\text{u})\!+\!H(V^n|W,A^n,X^n,Z^n,F)\nonumber\\
& \overset{(a)}{\leq} -H(U^n|Z^n,A^n) + H(W_\text{u},F_\text{u}|Z^n,A^n)\nonumber\\
&\;\quad + nH(V|A,X,Z)-H(W_\text{u} ,F_\text{u}|A^n,X^n,Z^n) \nonumber\\
&\;\quad -H(W_\text{v},F_\text{v}|A^n,X^n,Z^n,W_\text{u} ,F_\text{u}) +nH(U|A,X,Z,V)\nonumber\\
& \overset{(b)}{\leq}-H(U^n|Z^n,A^n) + H(W_\text{u}) +H(F_\text{u})\nonumber\\
&\;\quad + nH(U,V|A,X,Z)-H(U^n|A^n,X^n,Z^n)+n\epsilon^{\prime}_n \nonumber\\
&\;\quad -H(F_\text{v}|A^n,X^n,Z^n,W_\text{u} ,F_\text{u}) \nonumber\\
&\;\quad -H(W_\text{v}|A^n,X^n,Z^n,W_\text{u} ,F_\text{u},F_\text{v}) \nonumber\\
&  \overset{(c)}{\leq}n(-H(U|Z,A) + H(U|Y,A)+\epsilon)\nonumber\\
&\; \quad+ nH(V|U,A,X,Z)+n\epsilon^{\prime}_n -(H(F_\text{v})-\epsilon_n) \nonumber\\
&\;\quad -H(V^n|A^n,X^n,Z^n,U^n,F_\text{v})+H(V^n|A^n,X^n,Z^n,U^n)\nonumber\\
&\;\quad-I(V^n;W_\text{v},F_\text{v}|A^n,X^n,Z^n,U^n)\nonumber\\
&  \overset{(d)}{\leq}n(-H(U|Z,A) + H(U|Y,A)+\epsilon)\nonumber\\
&\; \quad +nH(V|U,A,X,Z) +n\epsilon^{\prime}_n-H(F_\text{v})+\epsilon_n\nonumber\\
&\;\quad -H(V^n|A^n,X^n,Z^n,U^n)+H(F_\text{v})\nonumber\\
&\;\quad +H(V^n|A^n,X^n,Z^n,U^n)\nonumber\\
&\;\quad\!-\!(H(W_\text{v}|A^n,X^n,Z^n,U^n)\!+\!H(F_\text{v}|A^n,X^n,Z^n,U^n)\!-\!\epsilon_n)\nonumber\\
&  \overset{(e)}{\leq}n(-H(U|Z,A) + H(U|Y,A)+\epsilon)\nonumber\\
&\;\quad + nH(V|U,A,X,Z)+n\epsilon_n^{\prime}+2\epsilon_n\nonumber\\
&\;\quad -(nR_\text{v}-2\epsilon_n)-(n\widetilde{R}_\text{v}-2\epsilon_n)\nonumber\\
& \overset{(f)}{=}n(I(U;Z|A) -I(U;Y|A)+\epsilon)\!+\!6\epsilon_n\nonumber\\
&\;\quad + n(I(V;Y|A,U)\!-\!I(V;X,Z|A,U)\!+\!\epsilon_n^{\prime}\!+\!\epsilon)\label{eq:case5step1}
\end{align}
where $(a)$ follows because $U^n$ determines $(W_\text{u},F_\text{u})$, $(b)$  follows by (\ref{eq:case5_1}) for some $\epsilon_n^{\prime}\rightarrow 0$ when $n\rightarrow\infty$, $(c)$ follows by (\ref{eq:R_utildechosen}), (\ref{eq:R_uchosen}), and (\ref{eq:independenceofFv}), and because $W_\text{u}$ and $F_\text{u}$ are determined by $U^n$, $(d)$ follows because $V^n$ determines $(F_\text{v}, W_\text{v})$, and $W_\text{v}$ and $F_\text{v}$ are almost independent given $(A^n,X^n,Z^n,U^n)$ by (\ref{eq:case5_3}), $(e)$ follows because $W_\text{v}$ and $F_\text{v}$ are almost independent of $(A^n,X^n,Z^n,U^n)$ and almost uniformly distributed by (\ref{eq:case5_3}), and $(f)$ follows by (\ref{eq:R_vtildechosen}) and (\ref{eq:R_vchosen}).

Combining (\ref{eq:achleakfirstpart}) and (\ref{eq:case5step1}), we obtain for Case 5
\begin{align}
&I(X^n;W,Z^n,F)\nonumber\\
&\; \leq  n\big(I(X;Z|A,U)+I(X;A,V,Y)-I(X;Y|A,U)  \big)\nonumber\\
&\quad +n\big(I(V;Z|A,X,U)\!-\!I(V;Y|A,X,U)\big)\nonumber\\
&\quad +n(I(V;Y|A,U)\!-\!I(V;X,Z|A,U)\!+\!\epsilon^{\prime\prime})\nonumber\\
&\;\quad +nI(U;Z|A) -nI(U;Y|A)\nonumber\\
&\;\overset{(a)}{=}  n\big(I(X,V;Z|A,U)+I(X;A,V,Y)-I(X;Y|A,V)  \big)\nonumber\\
&\;\quad -nI(V;Z|A,U) - nI(V;X|A,U,Z)+n\epsilon^{\prime\prime}\nonumber\\
&\;\quad +nI(U;Z|A) -nI(U;Y|A)\nonumber\\
&\;\overset{(b)}{=}  n\big(I(X;Z|A,V)+I(X;A,V,Y)-I(X;Y|A,V)  \big)\nonumber\\
&\;\quad - nI(V;X|A,U,Z)+n\epsilon^{\prime\prime}\nonumber\\
&\;\quad +nI(U;Z|A) -nI(U;Y|A)\nonumber\\
&\;\overset{(c)}{=}  n\big(I(X;A,U,Z)+I(U;Z|A) -I(U;Y|A)+\epsilon^{\prime\prime})\nonumber\\
\label{eq:case5privleak}
\end{align}
where $(a)$ follows from the Markov chain $U-V-(X,Y,A)$ and for some $\epsilon^{\prime\prime}>0$ such that $\epsilon^{\prime\prime}\rightarrow 0$ when $n\rightarrow\infty$, and $(b)$ and $(c)$ follow from the Markov chain $U-V-(X,Z,A)$.

\textbf{Case 6}: Suppose we have
\begin{align}
&R_\text{u}+\widetilde{R}_\text{u} \geq H(U|Z,A,X)\label{eq:case6_1}\\
&R_\text{u}+\widetilde{R}_\text{u} < H(U|Z,A) \label{eq:case6_2}\\
&R_\text{v}+\widetilde{R}_\text{v} \geq H(V|Z,A,X)\label{eq:case6_3}.
\end{align}
Then, we can recover $U^n$ from $(F_\text{u},W_\text{u},Z^n,A^n,X^n)$ by using a SW decoder \cite[Lemma 1]{OSRBAmin}, but $F_\text{u}$, $W_\text{u}$, and $(Z^n,A^n)$ are almost independent, and $F_\text{u}$ and $W_\text{u}$ are almost uniformly distributed by \cite[Theorem 1]{OSRBAmin}. Moreover, we can recover $V^n$ from $(F_\text{v},W_\text{v}, Z^n,A^n,X^n)$ by using a SW decoder \cite[Lemma 1]{OSRBAmin}. We obtain
\begin{align}
&-H(U^n|Z^n,A^n,W_\text{u},F_\text{u})\!+\!H(V^n|W,A^n,X^n,Z^n,F)\nonumber\\
&\;  \overset{(a)}{\leq} -H(U^n|Z^n,A^n) + H(W_\text{u},F_\text{u}|Z^n,A^n)\nonumber\\
&\;\qquad +H(V^n|W_\text{v},A^n,X^n,Z^n,F_\text{v})\nonumber\\
&\; \overset{(b)}{\leq}-H(U^n|Z^n,A^n) + H(W_\text{u}) +H(F_\text{u})+n\epsilon_n^{\prime}\nonumber\\
&\;   \overset{(c)}{\leq}n(-H(U|Z,A) + H(U|Y,A)+\epsilon+\epsilon_n^{\prime})\label{eq:case6step1}
\end{align}
where $(a)$ follows because $U^n$ determines $(F_\text{u},W_\text{u})$, $(b)$ follows by (\ref{eq:case6_3}) for some $\epsilon_n^{\prime}>0$ such that $\epsilon_n^{\prime}\rightarrow 0$ when $n\rightarrow\infty$, and $(c)$ follows by (\ref{eq:R_utildechosen}) and (\ref{eq:R_uchosen}).

Combining (\ref{eq:achleakfirstpart}) and (\ref{eq:case6step1}), we obtain for Case 6
\begin{align}
&I(X^n;W,Z^n,F)\nonumber\\
&\; \overset{(a)}{\leq}  n\big(I(X;Z|A,U)+I(X;A,V,Y)-I(X;Y|A,U)  \big)\nonumber\\
&\quad +n\big(I(V;Z|A,X,U)\!-\!I(V;Y|A,X,U)\big)\nonumber\\
&\quad +n(I(U;Z|A) -I(U;Y|A)+\epsilon^{\prime\prime})\nonumber\\
&\;\overset{(b)}{=} n\big(I(V,X;Z|A)+I(X;A,V,Y)-I(V,X;Y|A)  \big)\nonumber\\
&\quad +n\epsilon^{\prime\prime}\label{eq:case6privleak}
\end{align}
where $(a)$ follows for some $\epsilon^{\prime\prime}>0$ such that $\epsilon^{\prime\prime}\rightarrow 0$ when $n\rightarrow\infty$ and $(b)$ follows from the Markov chain $U-V-(A,X,Z,Y)$.

\textbf{Case 7}: Suppose we have
\begin{align}
&R_\text{u}+\widetilde{R}_\text{u} \geq H(U|Z,A) \label{eq:case7_1}\\
&R_\text{v}+\widetilde{R}_\text{v} \geq H(V|U,Z,A,X)\label{eq:case7_2}\\
&R_\text{v}+\widetilde{R}_\text{v} < H(V|Z,A,X)\label{eq:case7_3}.
\end{align}
Then, we can recover $U^n$ from $(F_\text{u},W_\text{u},Z^n,A^n)$ by using a SW decoder \cite[Lemma 1]{OSRBAmin}. Moreover, we can recover $V^n$ from $(F_\text{v},W_\text{v},U^n, Z^n,A^n,X^n)$ by using a SW decoder \cite[Lemma 1]{OSRBAmin}, but $F_\text{v}$, $W_\text{v}$, and $(Z^n,A^n,X^n)$ are almost independent, and $F_\text{v}$ and $W_\text{v}$ are almost uniformly distributed by \cite[Theorem 1]{OSRBAmin}. We obtain
\begin{align}
&-H(U^n|Z^n,A^n,W_\text{u},F_\text{u})\!+\!H(V^n|W,A^n,X^n,Z^n,F)\nonumber\\
&\; \overset{(a)}{\leq} H(V^n|W_\text{v},U^n,A^n,X^n,Z^n,F_\text{v})\nonumber\\
&\;\quad+H(U^n|W_\text{u},F_\text{u},A^n,Z^n)\nonumber\\
&\; \overset{(b)}{\leq}2n\epsilon_n^{\prime}\label{eq:case7step1}
\end{align}
where $(a)$ follows because $U^n$ determines $(F_\text{u},W_\text{u})$ and $(b)$ follows by (\ref{eq:case7_1}) and (\ref{eq:case7_2}) for some $\epsilon_n^{\prime}>0$ such that $\epsilon_n^{\prime}\rightarrow 0$ when $n\rightarrow\infty$.
 
 Combining (\ref{eq:achleakfirstpart}) and (\ref{eq:case7step1}), we obtain for Case 7
 \begin{align}
 &I(X^n;W,Z^n,F)\nonumber\\
 &\; \leq  n\big(I(X;Z|A,U)+I(X;A,V,Y)-I(X;Y|A,U)  \big)\nonumber\\
 &\quad +n\big(I(V;Z|A,X,U)\!-\!I(V;Y|A,X,U)\!+\!\epsilon^{\prime\prime}\big)\nonumber\\
  &\; \leq  n\big(I(V,X;Z|A,U)+I(X;A,V,Y) \big)\nonumber\\
  &\;\quad -nI(V,X;Y|A,U) +n\epsilon^{\prime\prime}\label{eq:case7privleak}
 \end{align}
 for some $\epsilon^{\prime\prime}>0$ such that $\epsilon^{\prime\prime}\rightarrow 0$ when $n\rightarrow\infty$.
 
\textbf{Case 8}: Suppose we have
\begin{align}
&R_\text{u}+\widetilde{R}_\text{u} \geq H(U|Z,A) \label{eq:case8_1}\\
&R_\text{v}+\widetilde{R}_\text{v} \geq H(V|Z,A,X)\label{eq:case8_2}.
\end{align}
Then, we can recover $U^n$ from $(F_\text{u},W_\text{u},Z^n,A^n)$ and $V^n$ from $(F_\text{v},W_\text{v}, Z^n,A^n,X^n)$ by using a SW decoder \cite[Lemma 1]{OSRBAmin}. We obtain
\begin{align}
&-H(U^n|Z^n,A^n,W_\text{u},F_\text{u})\!+\!H(V^n|W,A^n,X^n,Z^n,F)\nonumber\\
&\; \leq H(V^n|W_\text{v},A^n,X^n,Z^n,F_\text{v})\overset{(a)}{\leq}n\epsilon_n^{\prime}\label{eq:case8step1}
\end{align}
where $(a)$ follows by (\ref{eq:case8_2}) for some $\epsilon_n^{\prime}>0$ such that $\epsilon_n^{\prime}\rightarrow 0$ when $n\rightarrow\infty$.

Combining (\ref{eq:achleakfirstpart}) and (\ref{eq:case8step1}), we obtain for Case 8
\begin{align}
&I(X^n;W,Z^n,F)\nonumber\\
&\; \leq  n\big(I(X;Z|A,U)+I(X;A,V,Y)-I(X;Y|A,U)  \big)\nonumber\\
&\quad +n\big(I(V;Z|A,X,U)\!-\!I(V;Y|A,X,U)\!+\!\epsilon^{\prime\prime}\big)\nonumber\\
&\; \leq  n\big(I(V,X;Z|A,U)+I(X;A,V,Y) \big)\nonumber\\
&\;\quad -nI(V,X;Y|A,U) +n\epsilon^{\prime\prime}\label{eq:case8privleak}
\end{align}
for some $\epsilon^{\prime\prime}>0$ such that $\epsilon^{\prime\prime}\rightarrow 0$ when $n\rightarrow\infty$.

\textbf{Case 9}: Suppose we have
\begin{align}
&R_\text{u}+\widetilde{R}_\text{u} \geq H(U|Z,A) \label{eq:case9_1}\\
&R_\text{v}+\widetilde{R}_\text{v}< H(V|U,Z,A,X)\label{eq:case9_2}.
\end{align}
Then, we can recover $U^n$ from $(F_\text{u},W_\text{u},Z^n,A^n)$ by using a SW decoder \cite[Lemma 1]{OSRBAmin}. Moreover, $F_\text{v}$, $W_\text{v}$, and $(U^n,Z^n,A^n,X^n)$ are almost independent, and $F_\text{v}$ and $W_\text{v}$ are almost uniformly distributed by \cite[Theorem 1]{OSRBAmin}. We obtain
\begin{align}
&-H(U^n|Z^n,A^n,W_\text{u},F_\text{u})\!+\!H(V^n|W,A^n,X^n,Z^n,F)\nonumber\\
& \overset{(a)}{\leq} nH(V|A,X,Z)-H(W_\text{u} ,F_\text{u}|A^n,X^n,Z^n) \nonumber\\
&\;\quad -H(W_\text{v},F_\text{v}|A^n,X^n,Z^n,W_\text{u} ,F_\text{u}) +nH(U|A,X,Z,V)\nonumber\\
& \overset{(b)}{\leq} nH(U,V|A,X,Z)-H(U^n|A^n,X^n,Z^n)+n\epsilon^{\prime}_n \nonumber\\
&\;\quad -H(F_\text{v}|A^n,X^n,Z^n,W_\text{u} ,F_\text{u}) \nonumber\\
&\;\quad -H(W_\text{v}|A^n,X^n,Z^n,W_\text{u} ,F_\text{u},F_\text{v}) \nonumber\\
&  \overset{(c)}{\leq} nH(V|U,A,X,Z)+n\epsilon^{\prime}_n -(H(F_\text{v})-\epsilon_n) \nonumber\\
&\;\quad -H(V^n|A^n,X^n,Z^n,U^n,F_\text{v})+H(V^n|A^n,X^n,Z^n,U^n)\nonumber\\
&\;\quad-I(V^n;W_\text{v},F_\text{v}|A^n,X^n,Z^n,U^n)\nonumber\\
&  \overset{(d)}{\leq}nH(V|U,A,X,Z) +n\epsilon^{\prime}_n-H(F_\text{v})+\epsilon_n\nonumber\\
&\;\quad -H(V^n|A^n,X^n,Z^n,U^n)+H(F_\text{v})\nonumber\\
&\;\quad +H(V^n|A^n,X^n,Z^n,U^n)\nonumber\\
&\;\quad\!-\!(H(W_\text{v}|A^n,X^n,Z^n,U^n)\!+\!H(F_\text{v}|A^n,X^n,Z^n,U^n)\!-\!\epsilon_n)\nonumber\\
& \overset{(e)}{\leq}nH(V|U,A,X,Z)+n\epsilon_n^{\prime}+2\epsilon_n\nonumber\\
&\;\quad -(nR_\text{v}-2\epsilon_n)-(n\widetilde{R}_\text{v}-2\epsilon_n)\nonumber\\
& \overset{(f)}{=}n(I(V;Y|A,U)\!-\!I(V;X,Z|A,U)\!+\!\epsilon_n^{\prime}\!+\!\epsilon)\!+\!6\epsilon_n
\label{eq:case9step1}
\end{align}
where $(a)$ follows because $U^n$ determines $(W_\text{u},F_\text{u})$, $(b)$  follows by (\ref{eq:case9_1}) for some $\epsilon_n^{\prime}\rightarrow 0$ when $n\rightarrow\infty$, $(c)$ follows by (\ref{eq:independenceofFv}) and because $W_\text{u}$ and $F_\text{u}$ are determined by $U^n$, $(d)$ follows because $V^n$ determines $(F_\text{v}, W_\text{v})$, and $W_\text{v}$ and $F_\text{v}$ are almost independent given $(A^n,X^n,Z^n,U^n)$ by (\ref{eq:case9_2}), $(e)$ follows because $W_\text{v}$ and $F_\text{v}$ are almost independent of $(A^n,X^n,Z^n,U^n)$ and almost uniformly distributed by (\ref{eq:case9_2}), and $(f)$ follows by (\ref{eq:R_vtildechosen}) and (\ref{eq:R_vchosen}).

Combining (\ref{eq:achleakfirstpart}) and (\ref{eq:case9step1}), we obtain for Case 9
\begin{align}
&I(X^n;W,Z^n,F)\nonumber\\
&\; \overset{(a)}{\leq}  n\big(I(X;Z|A,U)+I(X;A,V,Y)-I(X;Y|A,U)  \big)\nonumber\\
&\quad +n\big(I(V;Z|A,X,U)\!-\!I(V;Y|A,X,U)\!+\!\epsilon^{\prime\prime}\big)\nonumber\\
&\quad +n(I(V;Y|A,U)\!-\!I(V;X,Z|A,U))\nonumber\\
&\; \overset{(b)}{=}  n(I(V;Z|A,U)+I(X;Z|A,V)+I(X;A,V,Y))\nonumber\\
&\quad -nI(V;Y|A,U) -nI(X;Y|A,V) +nI(V;Y|A,U)\nonumber\\
&\quad -nI(V;Z|A,U)-nI(V;X|A,U,Z)+n\epsilon^{\prime\prime}\nonumber\\
&\; \overset{(c)}{=}  nI(X;A,U,Z)+n\epsilon^{\prime\prime}
\label{eq:case9privleak}
\end{align}
where $(a)$ follows for some $\epsilon^{\prime\prime}>0$ such that $\epsilon^{\prime\prime}\rightarrow 0$ when $n\rightarrow\infty$ and $(b)$ follows from the Markov chain $U-V-(A,X,Z,Y)$.

Combining all cases and applying the selection lemma \cite[Lemma 2.2]{Blochbook}, there exists a binning that achieves all rate tuples $(R_\text{s},R_{\ell},R_\text{w},C)$ in the inner bound given in Theorem~\ref{theo:ActionBCgscs} for the key-leakage-storage-cost region $\mathcal{R}_{\text{gs}}$ for the GS model with strong secrecy when $n\rightarrow\infty$. 
\end{IEEEproof}
\subsection{Proof for the CS Model}
We use the achievability proof for the GS model. Suppose the key $S^{\prime}$, generated in the GS model together with the helper data $W^{\prime}=(W_\text{a}^{\prime}, W_\text{u}^{\prime}, W_\text{v}^{\prime})$ and public indices $F^{\prime}=(F_\text{a}^{\prime}, F_\text{u}^{\prime}, F_\text{v}^{\prime})$, have the same cardinality as an embedded secret key $S$, i.e., $|\mathcal{S}^{\prime}|=|\mathcal{S}|$, so that we achieve the same secret-key rate $R_\text{s}$ as in the GS model. The encoder $f_{2}(\cdot,\cdot)$ has inputs $(\widetilde{X}^n,S)$ and outputs $W=(S^{\prime}+S,W^{\prime})$. The decoder $g(\cdot,\cdot)$ has inputs $(W,Y^n)$ and output $\hat{S}=S^{\prime}+S-\hat{S}'$, where all addition and subtraction operations are modulo-$|\mathcal{S}|$. We use the decoder of the GS model to obtain $\hat{S}^{\prime}$. 

We have the error probability
\begin{align}
&\Pr[S\ne\hat{S}]=\Pr[S^{\prime}\ne\hat{S}^{\prime}]\label{eq:errorprobabilityachtheo2}
\end{align}
which is small due to the achievability proof for the GS model.

Using the one-time padding operation applied above, (\ref{eq:R_schosen}), and (\ref{eq:R_wchosen}), we can achieve a storage rate of
\begin{align}
&R_\text{w} \geq I(\widetilde{X};A,V)-I(U;Y|A)-I(V;Z|A,U)+4\epsilon
\end{align}
for the CS model.

Similar to the GS model, one can show that the expected cost constraint is satisfied with high probability by using the typical average lemma. 

We have the secrecy leakage of
\begin{align}
	&I(S;W,Z^n,F)\overset{(a)}{=} I(S;W,Z^n|F^{\prime})\nonumber\\
	&\; =I(S;W^{\prime},Z^n|F^{\prime})+I(S;S'+S,Z^n|W',F')\nonumber\\
	&\;\overset{(b)}{=} H(S^{\prime}+S,Z^n|W^{\prime},F^{\prime}) - H(S^{\prime},Z^n|W^{\prime},F^{\prime})\nonumber\\
	&\;= H(S^{\prime}+S|Z^n,W^{\prime},F^{\prime}) +H(Z^n|W^{\prime},F^{\prime})\nonumber\\
	&\;\quad- H(S^{\prime}|W^{\prime},F^{\prime})- H(Z^n|W^{\prime},F^{\prime},S^{\prime})\nonumber\\
	&\;\overset{(c)}{\leq} nR_\text{s}- H(S^{\prime}|W^{\prime},F^{\prime})+I(S';W^{\prime},Z^n|F^{\prime})\nonumber\\
	&\;\overset{(d)}{\leq} nR_\text{s}-(nR_\text{s}-2\epsilon_n)+I(S';W^{\prime},Z^n|F^{\prime})\overset{(e)}{\leq}3\epsilon_n
\end{align}
where \\
$(a)$ follows because $F=F^{\prime}$ and $S$ is independent of $F^{\prime}$;\\
$(b)$ follows because $S$ is independent of $(W^{\prime},F^{\prime},Z^n,S^{\prime})$; \\
$(c)$ follows because $|\mathcal{S}'|=|\mathcal{S}|$; \\
$(d)$ follows by (\ref{eq:secrecycond}) since $(S^{\prime},F_\text{v},W_\text{v})$ are almost mutually independent, uniformly distributed, and independent of  $(U^n,A^n,Z^n)$ so that $S^{\prime}$ is almost independent of $(F^{\prime},W^{\prime})$ and uniformly distributed;\\
 $(e)$ follows because the GS model satisfies the strong secrecy constraint (\ref{eq:secrecyleakage_cons}) by (\ref{eq:secrecycond}) for some $\epsilon_n>0$ such that $\epsilon_n\rightarrow0$ when $n\rightarrow\infty$.

We obtain the privacy-leakage of
\begin{align}
	&I(X^n;W,Z^n,F)\overset{(a)}{\leq}I(X^n;W,Z^n|F^{\prime})+3\epsilon_n\nonumber\\
	&\;\leq I(X^n;W^{\prime},Z^n|F^{\prime}) +H(S+S^{\prime}|Z^n,W^{\prime},F^{\prime})\nonumber\\
	&\qquad -H(S+S^{\prime}|Z^n,X^n,W^{\prime},F^{\prime},S^{\prime})+3\epsilon_n\nonumber\\
	&\overset{(b)}{\leq}  I(X^n;W^{\prime},Z^n|F^{\prime})+\log (|\mathcal{S}|)-H(S)+3\epsilon_n\nonumber\\
	&\overset{(c)}{=} I(X^n;W^{\prime},Z^n|F^{\prime}) +3\epsilon_n\label{eq:achcsprivleaknew}
\end{align}
where $(a)$ follows by (\ref{eq:XnandFareind}), $(b)$ follows because $S$ is independent of $(X^n,Z^n,W^{\prime},S^{\prime},F^{\prime})$ and $|\mathcal{S}'|=|\mathcal{S}|$, and $(c)$ follows from the uniformity of $S$. We therefore have the following results for nine different cases.

\textbf{Case 1}: Suppose (\ref{eq:case1_1}) and (\ref{eq:case1_2}). By combining (\ref{eq:case1privleak}) and (\ref{eq:achcsprivleaknew}), we obtain
\begin{align}
&I(X^n;W,Z^n,F)\nonumber\\
&\; \leq  n\big(I(V,X;Z|A)+I(X;A,V,Y)-I(V,X;Y|A)  \big)\nonumber\\
&\quad +n\epsilon^{(3)}\label{eq:case1privleakCS}
\end{align}
for some $\epsilon^{(3)}>0$ such that $\epsilon^{(3)}\rightarrow 0$ when $n\rightarrow\infty$.

\textbf{Case 2}: Suppose (\ref{eq:case2_1})-(\ref{eq:case2_3}). By combining (\ref{eq:case2privleak}) and (\ref{eq:achcsprivleaknew}), we have
\begin{align}
&I(X^n;W,Z^n,F)\nonumber\\
&\; \leq  n\big(I(V,X;Z|A,U)+I(X;A,V,Y)-I(V,X;Y|A,U)  \big)\nonumber\\
&\quad -nI(U;X|Z,A)\!+\!n\epsilon^{(3)}\label{eq:case2privleakCS}
\end{align}
for some $\epsilon^{(3)}>0$ such that $\epsilon^{(3)}\rightarrow 0$ when $n\rightarrow\infty$.

\textbf{Case 3}: Suppose (\ref{eq:case3_1}) and (\ref{eq:case3_2}). Combining (\ref{eq:case3privleak}) and (\ref{eq:achcsprivleaknew}), we have
\begin{align}
&I(X^n;W,Z^n,F)\leq  n(I(X;A,Z)+\epsilon^{(3)})\label{eq:case3privleakCS}
\end{align}
for some $\epsilon^{(3)}>0$ such that $\epsilon^{(3)}\rightarrow 0$ when $n\rightarrow\infty$.

\textbf{Case 4}: Suppose (\ref{eq:case4_1})-(\ref{eq:case4_4}). Combining (\ref{eq:case4privleak}) and (\ref{eq:achcsprivleaknew}), we obtain
\begin{align}
&I(X^n;W,Z^n,F)\nonumber\\
&\; \leq n\big(I(V,X;Z|A)+I(X;A,V,Y)-I(V,X;Y|A)  \big)\nonumber\\
&\quad +n\epsilon^{(3)}\label{eq:case4privleakCS}
\end{align}
for some $\epsilon^{(3)}>0$ such that $\epsilon^{(3)}\rightarrow 0$ when $n\rightarrow\infty$.
 
\textbf{Case 5}: Suppose (\ref{eq:case5_1})-(\ref{eq:case5_3}). By combining (\ref{eq:case5privleak}) and (\ref{eq:achcsprivleaknew}), we have
 \begin{align}
&I(X^n;W,Z^n,F)\nonumber\\
&\; \leq  n\big(I(X;A,U,Z)+I(U;Z|A) -I(U;Y|A)\!+\!\epsilon^{(3)}\big)\label{eq:case5privleakCS}
\end{align}
for some $\epsilon^{(3)}>0$ such that $\epsilon^{(3)}\rightarrow 0$ when $n\rightarrow\infty$.
 
\textbf{Case 6}: Suppose (\ref{eq:case6_1})-(\ref{eq:case6_3}). By combining (\ref{eq:case6privleak}) and (\ref{eq:achcsprivleaknew}), we obtain
\begin{align}
&I(X^n;W,Z^n,F)\nonumber\\
&\; \leq n\big(I(V,X;Z|A)+I(X;A,V,Y)-I(V,X;Y|A)  \big)\nonumber\\
&\quad\!+\!n\epsilon^{(3)}\label{eq:case6privleakCS}
\end{align}
for some $\epsilon^{(3)}>0$ such that $\epsilon^{(3)}\rightarrow 0$ when $n\rightarrow\infty$.

\textbf{Case 7}: Suppose (\ref{eq:case7_1})-(\ref{eq:case7_3}). Combining (\ref{eq:case7privleak}) and (\ref{eq:achcsprivleaknew}), we obtain
\begin{align}
&I(X^n;W,Z^n,F)\nonumber\\
&\; \leq n\big(I(V,X;Z|A,U)+I(X;A,V,Y)-I(V,X;Y|A,U) \big)\nonumber\\
&\;\quad+n\epsilon^{(3)}\label{eq:case7privleakCS}
\end{align}
for some $\epsilon^{(3)}>0$ such that $\epsilon^{(3)}\rightarrow 0$ when $n\rightarrow\infty$.

\textbf{Case 8}: Suppose (\ref{eq:case8_1}) and (\ref{eq:case8_2}). Combining (\ref{eq:case8privleak}) and (\ref{eq:achcsprivleaknew}), we obtain
\begin{align}
&I(X^n;W,Z^n,F)\nonumber\\
&\; \leq n\big(I(V,X;Z|A,U)+I(X;A,V,Y) -I(V,X;Y|A,U)\big)\nonumber\\
&\;\quad  +n\epsilon^{(3)}\label{eq:case8privleakCS}
\end{align}
for some $\epsilon^{(3)}>0$ such that $\epsilon^{(3)}\rightarrow 0$ when $n\rightarrow\infty$.

\textbf{Case 9}: Suppose (\ref{eq:case9_1}) and (\ref{eq:case9_2}). By combining (\ref{eq:case9privleak}) and (\ref{eq:achcsprivleaknew}), we obtain
\begin{align}
&I(X^n;W,Z^n,F) \leq nI(X;A,U,Z) +n\epsilon^{(3)}\label{eq:case9privleakCS}
\end{align}
for some $\epsilon^{(3)}>0$ such that $\epsilon^{(3)}\rightarrow 0$ when $n\rightarrow\infty$.

Using the selection lemma, there exists a binning that achieves all rate tuples $(R_\text{s},R_{\ell},R_\text{w},C)$ in the inner bound given in Theorem~\ref{theo:ActionBCgscs} for the key-leakage-storage-cost region $\mathcal{R}_{\text{cs}}$ for the CS model with strong secrecy when $n\rightarrow\infty$. 

\section{Outer Bounds for CLN Channels}\label{sec:OuterBoundProofs}
We use the following lemma, which is an extension of \cite[Lemma 1]{ChandraLessNoisy} proved for less-noisy BCs, to bound the privacy-leakage rate for CLN channels.
\begin{lemma}\label{lem:Chandraineq}
	\normalfont For a CLN channel $(X\geq Z|A,Y)$, we have
	\begin{align}
	&I(X_{i};X^{i-1}|W,S,A^n,Y^n) \nonumber\\
	&\qquad\geq I(X_{i};Z^{i-1}|W,S,A^n,Y^n),\label{eq:ChandraineqExtension1}\\
	&I(Z_{i};X^{i-1}|W,S,A^n,Y^n) \nonumber\\
    &\qquad\geq I(Z_{i};Z^{i-1}|W,S,A^n,Y^n)\label{eq:ChandraineqExtension2}
	\end{align} 
	for $i=1,2,\ldots,n$ if $(S,W)-(\widetilde{X}^n,Y^n,A^n)-(X^n,Z^n)$ form a Markov chain.
\end{lemma}
\begin{IEEEproof}
	Consider for any $1\leq j\leq i-1$ and $i=1,2,\ldots,n$ 
	\begin{align}
		&I(Z^{j-1},X_j^{i-1};X_i|W,S,A^n,Y^n)\nonumber\\
		&\;=I(Z^{j-1},X_{j+1}^{i-1};X_i|W,S,A^n,Y^n)\nonumber\\
		&\;\quad +I(X_j;X_i|W,S,A^n,Y^n,Z^{j-1},X_{j+1}^{i-1})\nonumber\\
		&\;\overset{(a)}{\geq}I(Z^{j-1},X_{j+1}^{i-1};X_i|W,S,A^n,Y^n)\nonumber\\
		&\;\quad +I(Z_j;X_i|W,S,A^n,Y^n,Z^{j-1},X_{j+1}^{i-1})\nonumber\\
		&\;=I(Z^{j},X_{j+1}^{i-1};X_i|W,S,A^n,Y^n)\label{eq:nairlemmaproofpart1}
	\end{align}
	where $X_{i}^{i-1}$ and $Z^{0}$ are considered to be constant and $(a)$ follows from the inequality
	\begin{align}
		&I(X_i;X_j|Y_j,A_j,W,S,Z^{j-1},X_{j+1}^{i-1},Y^{n\setminus j},A^{n\setminus j})\nonumber\\
		&\overset{(a.1)}{\geq}I(X_i;Z_j|Y_j,A_j,W,S,Z^{j-1},X_{j+1}^{i-1},Y^{n\setminus j},A^{n\setminus j})\label{eq:nairlemmaproofpart2}
	\end{align}
	where $Y^{{n\setminus j}}$ is the set of random variables $\{Y_1,Y_2,\ldots,Y_{j-1},Y_{j+1},\ldots,Y_n\}$ and $(a.1)$ follows for a CLN channel such that $(X\!\geq\! Z|A,Y)$ by (\ref{eq:condforCLN}) and since $(W,S,Z^{j-1},X_{j+1}^{i-1},Y^{n \setminus j},A^{n \setminus j},X_i)\!-\!(A_j,\widetilde{X}_j,Y_j)\!-\!(X_j,Z_j)$ form a Markov chain. Apply (\ref{eq:nairlemmaproofpart1}) repetitively for $j=1,2,\ldots,i-1$ such that
	\begin{align}
		&I(X^{i-1};X_i|W,S,A^n,Y^n)\nonumber\\
		&\;\geq I(Z_{1},X_2^{i-1};X_i|W,S,A^n,Y^n)\nonumber\\
		&\;\geq I(Z^{2},X_3^{i-1};X_i|W,S,A^n,Y^n)\nonumber\\
		&\;\geq \ldots\geq I(Z^{i-1};X_i|W,S,A^n,Y^n)\label{eq:nairlemmaproofpart3}
	\end{align} 
	which is the proof for (\ref{eq:ChandraineqExtension1}). The proof of (\ref{eq:ChandraineqExtension2}) follows by replacing $X_i$ with $Z_i$ in (\ref{eq:nairlemmaproofpart1}), (\ref{eq:nairlemmaproofpart2}), and (\ref{eq:nairlemmaproofpart3}) since $(W,S,Z^{j-1},X_{j+1}^{i-1},Y^{n \setminus j},A^{n \setminus j},Z_i)\!-\!(A_j,\widetilde{X}_j,Y_j)\!-\!(X_j,Z_j)$ also form a Markov chain.
\end{IEEEproof}

\subsection{Proofs of Outer Bounds}
Suppose for some $\delta_n\!>\!0$ and $n\geq 1$, there is a pair of encoders and decoders such that (\ref{eq:reliability_cons})-(\ref{eq:cost_cons}) are satisfied for all CLN channels such that $(X\geq Z|A,Y)$ and $(Z\geq Y|A,X)$ by some key-leakage-storage-cost tuple $(R_{\text{s}}, R_\ell,R_{\text{w}},C)$. If $(X\geq Z|A,Y)$ and $(Z\geq Y|A,X)$, we also have  $(X\geq Y|A,Z)$. Using (\ref{eq:reliability_cons}) and Fano's inequality, we obtain
\begin{align}
H(S|W,Y^n)\!\overset{(a)}{\leq}\!H(S|\hat{S})\!\leq\!n\epsilon_n \label{eq:fanoapp} 
\end{align}
where $(a)$ permits randomized decoding, $\epsilon_n\!=\!\delta_n R_{\text{s}} \!+\!H_b(\delta_n)/n$, where $H_b(\delta) = -\delta\log \delta - (1-\delta)\log(1-\delta)$ is the binary entropy function, and $\epsilon_n\!\rightarrow\!0$ if $\delta_n\!\rightarrow\!0$.

Let $U_{i}\triangleq (W,A^{n \setminus i},Y_{i+1}^{n},Z^{i-1})$ and $V_{i}\triangleq (S,W,A^{n \setminus i},Y_{i+1}^n,Z^{i-1})$, which satisfy the Markov chain $U_i-V_i-(A_i,\widetilde{X}_i)-(A_i,\widetilde{X}_i,X_i)-(Y_i,Z_i)$ for all $i=1,2,\ldots,n$.

\emph{Secret-key Rate}: We obtain for the GS and CS models
\begin{align}
&n(R_\text{s}-\delta_n) \overset{(a)}{\leq} H(S)\overset{(b)}{\leq} H(S|W,Z^n) +\delta_n\nonumber\\
& \;\overset{(c)}{=}H(S|W,A^n,Z^n)+\delta_n\nonumber\\
& \;\overset{(d)}{\leq}H(S|W,A^n,Z^n) -H(S|W,A^n,Y^n) + n\epsilon_n+\delta_n\nonumber\\
&\;=\sum_{i=1}^n \big[I(S;Y_i|W,A^n,Y_{i+1}^n) - I(S;Z_i|W,A^n,Z^{i-1})\big]\nonumber\\
&\;\qquad+ n\epsilon_n+\delta_n\nonumber\\
&\;\overset{(e)}{=} \sum_{i=1}^n \Big[I(S;Y_i|W,A^n,Y_{i+1}^n,Z^{i-1}) \nonumber\\
&\;\qquad- I(S;Z_i|W,A^n,Y_{i+1}^n,Z^{i-1})\Big] + n\epsilon_n+\delta_n\nonumber\\
&\;\overset{(f)}{=}\! \sum_{i=1}^n \big[ I(V_i;Y_i|A_i,U_i) \!-\! I(V_i;Z_i|A_i,U_i)\!+\! \epsilon_n\big]\!+\!\delta_n\label{eq:keyrateouterproof}
\end{align}
where $(a)$ follows by (\ref{eq:uniformity_cons}), $(b)$ follows by (\ref{eq:secrecyleakage_cons}), $(c)$ follows from the deterministic action encoder, $(d)$ follows by (\ref{eq:fanoapp}), $(e)$ follows from Csisz\'{a}r's sum identity \cite{CsiszarKornerbook2011}, and $(f)$ follows from the definitions of $U_i$ and $V_i$.

\emph{Storage Rate}: We obtain for the GS model
\begin{align}
&n(R_\text{w}+\delta_n) \overset{(a)}{\geq} \log|\mathcal{W}| \geq H(W)\nonumber\\
&\overset{(b)}{=} H(A^{n}) + H(W|A^{n})\nonumber\\
&\geq [H(A^{n})-H(A^{n}|\widetilde{X}^{n},Z^{n})]+ [H(W|A^{n},Y^{n}) \nonumber\\
&\qquad -H(W|A^{n},\widetilde{X}^{n},Y^{n},Z^{n}) ] \nonumber\\
&\!=\!H(\widetilde{X}^{n},Z^{n})\!-\!H(\widetilde{X}^{n},Z^{n}|A^{n})\!+\!H(\widetilde{X}^{n},Z^{n}|A^{n},Y^{n})\nonumber\\
&\qquad-H(\widetilde{X}^{n},Z^{n}|A^{n},Y^{n},W)\nonumber\\
&\!=\! H(\widetilde{X}^{n})\!+\! H(Z^{n}|\widetilde{X}^{n})\!-\!H(Y^{n}|A^{n})\nonumber\\
&\qquad \!+\!H(Y^{n},Z^{n}|\widetilde{X}^{n},A^{n})-H(Z^{n}|\widetilde{X}^{n},A^{n})\nonumber\\
&\qquad -H(\widetilde{X}^{n},Z^{n}|A^{n},Y^{n},W,S)\nonumber\\
&\qquad-I(\widetilde{X}^{n},Z^n;S|A^n,Y^n,W)\nonumber\\
&\!\geq\!   H(\widetilde{X}^{n}) \!-\!H(Y^{n}|A^{n})\!+\!H(Y^{n},Z^{n}|\widetilde{X}^{n},A^{n})\nonumber\\
&\qquad \!-\!H(\widetilde{X}^{n},Z^{n}|A^{n},Y^{n},W,S)\!-\!H(S|A^n,Y^n,W)\nonumber\\
&\overset{(c)}{\geq}  \sum_{i=1}^{n} \Big[H(\widetilde{X}_{i}) - H(Y_i|A_{i})+H(Y_i,Z_i|\widetilde{X}_{i},A_{i}) \nonumber\\
&\qquad-H(\widetilde{X}_{i},Z_i|A^{n},Y^{n},W,S,\widetilde{X}^{i-1},Z^{i-1})\Big]-n\epsilon_n\nonumber\\
&\overset{(d)}{\geq} \sum_{i=1}^{n} \Big[H(\widetilde{X}_{i}) - H(Y_i|A_{i})+H(Y_i|\widetilde{X}_{i},A_{i},Z_i) \nonumber\\
&\qquad+H(Z_i|\widetilde{X}_{i},A_{i})-H(\widetilde{X}_{i},Z_i|A_{i},Y_i,V_{i})\Big]-n\epsilon_n\nonumber\\
&\geq \sum_{i=1}^{n} \Big[I(\widetilde{X}_{i};A_{i})+I(V_{i};\widetilde{X}_{i}|A_{i},Y_i)-\epsilon_n\Big]\label{eq:storageGSouterproof}
\end{align}
where $(a)$ follows by (\ref{eq:storage_cons}), $(b)$ follows from the deterministic action encoder, $(c)$ follows by (\ref{eq:fanoapp}), and $(d)$ follows from the definition of $V_i$. 

We obtain for the CS model
\begin{align}
&n(R_\text{w}+\delta_n) \overset{(a)}{\geq} \log|\mathcal{W}| \geq H(W)\nonumber\\
&\overset{(b)}{=} H(A^{n}) + H(W|A^{n})\nonumber\\
&\overset{(c)}{\geq} H(A^{n})-H(A^{n}|\widetilde{X}^{n},Z^{n})+H(A^{n}|\widetilde{X}^{n},Z^{n})\nonumber\\
&\qquad+ H(W|A^{n},Y^{n})  -H(W|A^{n},\widetilde{X}^{n},Y^{n},Z^{n})\nonumber\\
&\qquad +H(W|A^{n},\widetilde{X}^{n})\nonumber\\
&=H(\widetilde{X}^{n},Z^{n})-H(\widetilde{X}^{n},Z^{n}|A^{n})+H(A^{n}|\widetilde{X}^{n},Z^{n})\nonumber\\
&\qquad+H(\widetilde{X}^{n},Z^{n}|A^{n},Y^{n})-H(\widetilde{X}^{n},Z^{n}|A^{n},Y^{n},W)\nonumber\\
&\qquad+H(W|A^{n},\widetilde{X}^{n})\nonumber\\
&= H(\widetilde{X}^{n}) + H(Z^{n}|\widetilde{X}^{n})-H(Y^{n}|A^{n})\nonumber\\
&\qquad +H(Y^{n},Z^{n}|\widetilde{X}^{n},A^{n})-H(Z^{n}|\widetilde{X}^{n},A^{n})\nonumber\\
&\qquad+H(A^{n}|\widetilde{X}^{n},Z^{n})-H(\widetilde{X}^{n},Z^{n}|A^{n},Y^{n},W,S)\nonumber\\
&\qquad -I(\widetilde{X}^{n},Z^n;S|A^n,Y^n,W)+H(W|A^{n},\widetilde{X}^{n})\nonumber\\
&=   H(\widetilde{X}^{n}) + I(Z^n;A^n|\widetilde{X}^{n})-H(Y^{n}|A^{n})\nonumber\\
&\qquad+H(Y^{n},Z^{n}|\widetilde{X}^{n},A^{n})+H(A^{n}|\widetilde{X}^{n},Z^{n})\nonumber\\
&\qquad-H(\widetilde{X}^{n},Z^{n}|A^{n},Y^{n},W,S)-H(S|A^n,Y^n,W)\nonumber\\
&\qquad + H(S|A^n,Y^n,W,\widetilde{X}^{n},Z^n)+H(W|A^{n},\widetilde{X}^{n})\nonumber\\
&\overset{(d)}{=}\!  H(\widetilde{X}^{n})\!+\!H(W,A^n,S|\widetilde{X}^{n})-H(Y^{n}|A^{n})\nonumber\\
&\qquad\!+\!H(Y^{n},Z^{n}|\widetilde{X}^{n},A^{n})\!-\!H(\widetilde{X}^{n},Z^{n}|A^{n},Y^{n},W,S)\nonumber\\
&\qquad -H(S|A^n,Y^n,W)\nonumber\\
&\overset{(e)}{\geq}\!  H(\widetilde{X}^{n})\!+\!H(S)\!-\!H(Y^{n}|A^{n})\!+\!H(Y^{n},Z^{n}|\widetilde{X}^{n},A^{n})\nonumber\\
&\qquad -H(\widetilde{X}^{n},Z^{n}|A^{n},Y^{n},W,S)-H(S|A^n,Y^n,W)\nonumber\\
&\geq  H(\widetilde{X}^{n})-H(Y^{n}|A^{n})+H(Y^{n},Z^{n}|\widetilde{X}^{n},A^{n})\nonumber\\
&\qquad -H(\widetilde{X}^{n},Z^{n}|A^{n},Y^{n},W,S)\nonumber\\
&\qquad+H(S|A^n,Z^n,W)-H(S|A^n,Y^n,W)\nonumber\\
&\overset{(f)}{=} \sum_{i=1}^{n} \Big[H(\widetilde{X}_{i}) - H(Y_i|A_{i})+H(Y_i,Z_i|\widetilde{X}_{i},A_{i})\nonumber\\
&\qquad-H(\widetilde{X}_{i},Z_i|A^{n},Y^{n},W,S,\widetilde{X}^{i-1},Z^{i-1})\nonumber\\
&\qquad+ I(S;Y_i|W,A^n,Y_{i+1}^n) - I(S;Z_i|W,A^n,Z^{i-1})\Big]\nonumber\\
&\overset{(g)}{=}  \sum_{i=1}^{n} \Big[H(\widetilde{X}_{i}) - H(Y_i|A_{i})+H(Y_i,Z_i|\widetilde{X}_{i},A_{i})\nonumber\\
&\qquad-H(\widetilde{X}_{i},Z_i|A^{n},Y^{n},W,S,\widetilde{X}^{i-1},Z^{i-1})\nonumber\\
&\qquad+ I(S;Y_i|W,A^n,Y_{i+1}^n,Z^{i-1})\nonumber\\
&\qquad- I(S;Z_i|W,A^n,Y_{i+1}^n,Z^{i-1})\Big]\nonumber\\
&\overset{(h)}{\geq} \sum_{i=1}^{n} \Big[H(\widetilde{X}_{i}) - H(Y_i|A_{i})+H(Y_i|\widetilde{X}_{i},A_{i},Z_i)\nonumber\\
&\qquad+H(Z_i|\widetilde{X}_{i},A_{i})-H(\widetilde{X}_{i},Z_i|A_{i},Y_i,V_{i})\nonumber\\
&\qquad+I(V_i;Y_i|A_i,U_i) - I(V_i;Z_i|A_i,U_i)\Big]\nonumber\\
&\geq \sum_{i=1}^{n}\Big[ I(\widetilde{X}_{i};A_{i})+I(V_{i};\widetilde{X}_{i}|Y_i,A_{i})+I(V_i;Y_i|A_i,U_i)\nonumber\\
&\qquad - I(V_i;Z_i|A_i,U_i)\Big]\nonumber
\end{align}
\begin{align}
&\overset{(i)}{=}\sum_{i=1}^{n} \Big[I(\widetilde{X}_{i};A_{i},V_{i}) - I(U_i;Y_i|A_i) \nonumber\\
&\qquad- I(V_i;Z_i|A_i,U_i)\Big]\label{eq:storageCSouterproof}
\end{align}
where $(a)$ follows by (\ref{eq:storage_cons}), $(b)$ follows from the deterministic action encoder, $(c)$ follows from the Markov chain $W-(A^n,\widetilde{X}^n)-(Y^n,Z^n)$, $(d)$ follows from the Markov chain $S-(A^n,\widetilde{X}^n,W)-(Y^n,Z^n)$, $(e)$ follows because the embedded key $S$ is independent of $\widetilde{X}^n$, $(f)$ follows when $Y_{n+1}^n$ and $Z^{0}$ are considered to be constant, $(g)$ follows from Csisz\'{a}r's sum identity. We use the definitions of $U_i$ and $V_i$ in $(h)$, and $(i)$ follows because $U_i-V_i-(A_i,\widetilde{X}_i)-(Y_i,Z_i)$ form a Markov chain for all $i=1,2,\ldots,n$. 

\emph{Privacy-leakage Rate}: We have for GS and CS models that
\begin{align}
&n(R_\ell + \delta_n) \overset{(a)}{\geq} I(X^{n};W,A,Z^{n})\nonumber\\
&=  H(X^n)\!-\!H(X^{n}|W,S,A^{n},Y^{n}) \!-\!I(X^n;S|W,A^n,Y^n)\nonumber\\
&\;\qquad-I(X^{n};Y^{n}|W,A^{n})+I(X^{n};Z^{n}|W,A^{n})\nonumber\\
& \overset{(b)}{\geq} \sum_{i=1}^{n} \Big[H(X_i)-H(X_{i}|W,S,A^{n},Y^{n},X^{i-1}) -\epsilon_n \nonumber\\
&\;\qquad \!-\!H(Y_i|W,A^{n},Y_{i+1}^{n})\!+\! H(Y_i|X_{i},A_{i})\nonumber\\
&\;\qquad +H(Z_i|W,A^{n},Z^{i-1})-H(Z_i|X_{i},A_{i})\Big]\nonumber\\
&\;\qquad \!-\!I(Y^n;W|A^n,X^n) +I(Z^n;W|A^n,X^n)\nonumber\\
& \overset{(c)}{\geq} \sum_{i=1}^{n}\Big[ H(X_i)-H(X_{i}|W,S,A^{n},Y^{n},Z^{i-1}) \nonumber\\
&\;\qquad -I(X_{i};Y_i|A_{i})+ H(Y_i|A_{i}) +I(X_{i};Z_i|A_{i})\nonumber\\
 &\;\qquad - H(Z_i|A_{i}) -H(Y_i|W,A^{n},Y_{i+1}^{n})\nonumber\\
 &\;\qquad+H(Z_i|W,A^{n},Z^{i-1})-\epsilon_n\Big]\nonumber\\
 &\;\qquad -I(Y^n;W|A^n,X^n) +I(Z^n;W|A^n,X^n)\nonumber\\
& \overset{(d)}{\geq} \sum_{i=1}^{n} \Big[H(X_i)-H(X_{i}|V_{i},A_{i},Y_i)-I(X_{i};Y_i|A_{i})\nonumber\\
&\;\qquad + I(W,Y_{i+1}^{n},A^{n \setminus i};Y_i|A_{i})+I(X_{i};Z_i|A_{i})\nonumber\\
 & \;\qquad -I(W,Z^{i-1},A^{n \setminus i};Z_i|A_{i})\!-\!\epsilon_n\Big]\nonumber\\
  &\;\qquad -I(Y^n;W|A^n,X^n) +I(Z^n;W|A^n,X^n)\nonumber\\
&\overset{(e)}{=} \sum_{i=1}^{n} \Big[I(X_{i};A_{i},V_{i},Y_i) \!-\!I(X_{i};Y_i|A_{i})\!+\!I(X_{i};Z_i|A_{i})\!-\!\epsilon_n \nonumber\\
&\;\qquad+ I(W,Y_{i+1}^{n},Z^{i-1},A^{n \setminus i};Y_i|A_{i})\nonumber\\
&\;\qquad-I(W,Z^{i-1},Y_{i+1}^n,A^{n \setminus i};Z_i|A_{i})\Big] \nonumber\\
&\;\qquad -I(Y^n;W|A^n,X^n) +I(Z^n;W|A^n,X^n)\nonumber\\
&\overset{(f)}{=} \sum_{i=1}^{n} \Big[I(X_{i};A_{i},V_{i},Y_i) \!-\!I(X_{i};Y_i|A_{i})\!+\!I(X_{i};Z_i|A_{i})\!-\!\epsilon_n \nonumber\\
&\;\qquad+ I(U_i;Y_i|A_{i})-I(U_i;Z_i|A_{i})\Big] \nonumber\\
&\;\qquad -I(Y^n;W|A^n,X^n) +I(Z^n;W|A^n,X^n)\label{eq:privleakouterpart1}
\end{align}
where $(a)$ follows by (\ref{eq:leakage_cons}) and from the deterministic action encoder, $(b)$ follows by (\ref{eq:fanoapp}), $(c)$ follows by (\ref{eq:ChandraineqExtension1}) for CLN channels $(X\geq Z|A,Y)$, $(d)$ follows from the definition of $V_{i}$ and  the deterministic action encoder, $(e)$ follows from Csisz\'{a}r's sum identity, and $(f)$ follows from the definition of $U_i$.

Consider the extra terms  
\begin{align}
	&-I(Y^n;W|A^n,X^n) +I(Z^n;W|A^n,X^n) \nonumber\\
	&\qquad= H(W|A^n,X^n,Y^n)-H(W|A^n,X^n,Z^n) \label{eq:extraterms}
\end{align}
in (\ref{eq:privleakouterpart1}). We cannot apply Csisz\'{a}r's sum identity to the terms in (\ref{eq:extraterms}) due to the conditioning on $X^n$, so one should find another method to have a single-letter expression for  (\ref{eq:extraterms}). We use the following lemma to replace the terms in (\ref{eq:extraterms}) with single-letter expressions and make further assumptions to bound the extra term.

\begin{lemma}\label{lem:iidexists}
	Consider the model given in Fig.~\ref{fig:hidden}. There exists a random variable $\widebar{W}$ such that $(\widebar{W}^n,A^n,\widetilde{X}^n,X^n,Y^n,Z^n)$ are i.i.d., $\widebar{W}-(A,\widetilde{X})-(A,\widetilde{X},X) - (Y,Z)$ form a Markov chain, and
	\begin{align}
		&H(W|A^n,X^n,Y^n)-H(W|A^n,X^n,Z^n)\nonumber\\
		&\; = n\Big(H(\widebar{W}|A,X,Y)-H(\widebar{W}|A,X,Z)\Big)\label{eq:lemmaequality}
	\end{align}
	when $n\rightarrow\infty$. 
\end{lemma}
\begin{IEEEproof}[Proof Sketch]
	Consider the encoder $f_\text{s}(\cdot)$ and decoder $g_\text{s}(\cdot)$ of a lossless source code such that $W=f_\text{s}(\widebar{W}^n)$ and $\widehat{\widebar{W}}^n=g(W)$, where $\widehat{\widebar{W}}^n$ is an estimate of $\widebar{W}^n$. Suppose the lossless source code achieves the optimal compression rate of $R=H(\widebar{W}) = \frac{H(W)}{n}$ when $n\rightarrow\infty$. We thus obtain
	\begin{align}
		&H(W|\widebar{W}^n) = H(W) - nH(\widebar{W}) + H(\widebar{W}^n|W)\nonumber\\
		&\; = H(\widebar{W}^n|W)\overset{(a)}{\leq} H(\widebar{W}^n|\widehat{\widebar{W}}^n)\overset{(b)}{\leq}n\epsilon_n \label{eq:proofforlemmaiidexistpart1}
	\end{align}
	where $(a)$ follows from the data processing inequality applied to the Markov chain $\widebar{W}^n\!-\!W\!-\!\widehat{\widebar{W}}^n$ and $(b)$ follows from Fano's inequality for some $\epsilon_n>0$ such that $\epsilon_n\rightarrow0$ when $n\rightarrow\infty$. Since $\widehat{\widebar{W}}^n\!-\!W\!-\!(A^n,\widetilde{X}^n)\!-\!(A^n,\widetilde{X}^n,X^n) \!-\! (Y^n,Z^n)$ form a Markov chain, the proof of Lemma~\ref{lem:iidexists} follows by (\ref{eq:proofforlemmaiidexistpart1}). 
\end{IEEEproof}

Suppose the joint probability distribution $P_{X\widetilde{X}AYZ}$ satisfies also the condition to be a CLN channel $(Z\geq Y|A,X)$. For the random variable $\widebar{W}$ defined in Lemma~\ref{lem:iidexists}, we then have
\begin{align}
	H(\widebar{W}|A,X,Y)\geq H(\widebar{W}|A,X,Z).\label{eq:secondCLNcond}
\end{align}
Thus, by combining (\ref{eq:privleakouterpart1}), (\ref{eq:extraterms}), (\ref{eq:lemmaequality}), and (\ref{eq:secondCLNcond}) for  a CLN channel $(Z\geq Y|A,X)$, we obtain

\begin{align}
&n(R_\ell + \delta_n)  \nonumber\\
&\;\geq \sum_{i=1}^{n} \Big[I(X_{i};A_{i},V_{i},Y_i) \!-\!I(X_{i};Y_i|A_{i})\!+\!I(X_{i};Z_i|A_{i})\!-\!\epsilon_n \nonumber\\
&\;\qquad+ I(U_i;Y_i|A_{i})-I(U_i;Z_i|A_{i})\Big] \label{eq:privleakouterproof}
\end{align}
when $n\rightarrow\infty$.

\emph{Expected Action Cost}: We have
\begin{align}
C+\delta_n \overset{(a)}{\geq} \mathbb{E}\big[\Gamma^{(n)}(A^{n})\big] =\frac{1}{n}\sum_{i=1}^{n} \mathbb{E}\big[\Gamma(A_{i})\big]\label{eq:actioncostouterproof}
\end{align}
where $(a)$ follows by (\ref{eq:cost_cons}).

Introduce a uniformly distributed time-sharing random variable $\displaystyle Q\!\sim\! \text{Unif}[1\!:\!n]$ independent of other random variables. Define $X\!=\!X_Q$, $\displaystyle \widetilde{X}\!=\!\widetilde{X}_Q$, $\displaystyle Y\!=\!Y_Q$, $\displaystyle Z\!=\!Z_Q$, $\displaystyle A\!=\!A_Q$, $U\!=\!(U_Q,\!Q)$, and $V\!=\!(V_Q,\!Q)$  so that $\displaystyle U\!-V-(A,\widetilde{X})-(A,\widetilde{X},X)-(Y,Z)$ form a Markov chain. The outer bound for all CLN channels such that $(X\geq Z|A,Y)$ and $(Z\geq Y|A,X)$ for the GS model follows by using the introduced random variables in (\ref{eq:keyrateouterproof}), (\ref{eq:storageGSouterproof}), (\ref{eq:privleakouterproof}), and (\ref{eq:actioncostouterproof}), and letting $\delta_n\rightarrow0$. Similarly, the outer bound for the same class of channels for the CS model follows by using the introduced random variables in (\ref{eq:keyrateouterproof}), (\ref{eq:storageCSouterproof}), (\ref{eq:privleakouterproof}), and (\ref{eq:actioncostouterproof}), and letting $\delta_n\rightarrow0$.

\textit{Cardinality Bounds}: We use the support lemma \cite[Lemma 15.4]{CsiszarKornerbook2011}. The bound in (\ref{eq:CSRwOuter}) can be written as the sum of the bounds in (\ref{eq:GSRsOuter}) and (\ref{eq:GSRwOuter}). Therefore, the same cardinality bounds can be used for the outer bounds of the GS and CS models. One can preserve $P_{A\widetilde{X}}$ by using $|\mathcal{A}||\mathcal{\widetilde{X}}|-1$ real-valued continuous functions. We have to preserve four more expressions, i.e., $H(V|A,U,Z) - H(V|A,U,Y)$, $H(X|U,V,A,Y)$, $H(\widetilde{X}|U,V,A,Y)$, and $H(Y|A,U)-H(Z|A,U)$. Thus, one can limit the cardinality $|\mathcal{U}|$ of $U$ to   $|\mathcal{U}|\leq|\mathcal{A}||\mathcal{\widetilde{X}}|+3$. Similarly, in addition to the  $|\mathcal{A}||\mathcal{\widetilde{X}}|-1$ real-valued continuous functions, one should preserve three more expressions, i.e., $H(X|A,V,Y)$, $H(\widetilde{X}|A,V,Y)$, and $H(Y|A,U,V)-H(Z|A,U,V)$ for the auxiliary random variable $V$. Furthermore, to satisfy the Markov condition $U-V-(A,\widetilde{X})-(A,\widetilde{X},X)-(Y,Z)$, one can limit the cardinality $|\mathcal{V}|$ of $V$ to $|\mathcal{V}|\leq(|\mathcal{A}||\mathcal{\widetilde{X}}|+3)(|\mathcal{A}||\mathcal{\widetilde{X}}|+2)$.

\section{Conclusion}\label{sec:conclusion}
We derived inner and outer bounds for the key-leakage-storage-cost regions for a hidden (noisy) identifier source with correlated noise components at the encoder and decoder to generate or embed secrets when a cost-constrained action sequence controls the decoder measurements. The correlation between the noise components is provided by a model where the encoder measurement is an input to the decoder measurement channel, as an extension of a BC model. Side information at the eavesdropper that is correlated with the encoder and decoder measurements is also considered since it is a realistic assumption for biometric identifiers. The achievability proofs of the inner bounds involve a random encoding step using the OSRB method that provides strong secrecy. The main difference between the bounds for the GS and CS models is the increased storage rate for the CS model as compared to the GS model. The outer bounds are given for CLN channels, for which important inequalities are derived. The inner and outer bound terms match for the secret-key rate, storage rate, and cost, and are different for the privacy-leakage rate.

We illustrated achievable cost vs. secret-key rate pairs for a set of storage rates with an example, where source and channel parameters were motivated by realistic authentication scenarios that use RO PUFs. We showed that an action sequence significantly decreases the necessity of reliable measurement channels to achieve the maximum secret-key rate. This reduction in the required reliability allows to have a larger hardware area available for public storage, which is illustrated to further increase the secret-key rate achieved for the same expected action cost. 
 
In future work, we will study the possibility of introducing a third auxiliary random variable as it might be possible to find single-letter expressions for the extra multi-letter terms in the outer bounds by defining another auxiliary random variable.

\section*{Acknowledgment}
O. G\"unl\"u and R. F. Schaefer were supported by the German Federal Ministry of Education and Research (BMBF) within the national initiative for ``Post Shannon Communication (NewCom)'' under the Grant 16KIS1004. The work of H. V. Poor was supported by the U.S. National Science Foundation under Grants CCF-0939370, CCF-1513915, and CCF-1908308. O. G\"unl\"u thanks Gerhard Kramer for his previous suggestions that indirectly helped to obtain Lemma~\ref{lem:iidexists} and Matthieu Bloch for his insightful comments.

\bibliographystyle{IEEEtran}
\bibliography{IEEEabrv,references}

\begin{thebibliography}{10}
\providecommand{\url}[1]{#1}
\csname url@samestyle\endcsname
\providecommand{\newblock}{\relax}
\providecommand{\bibinfo}[2]{#2}
\providecommand{\BIBentrySTDinterwordspacing}{\spaceskip=0pt\relax}
\providecommand{\BIBentryALTinterwordstretchfactor}{4}
\providecommand{\BIBentryALTinterwordspacing}{\spaceskip=\fontdimen2\font plus
\BIBentryALTinterwordstretchfactor\fontdimen3\font minus
  \fontdimen4\font\relax}
\providecommand{\BIBforeignlanguage}[2]{{%
\expandafter\ifx\csname l@#1\endcsname\relax
\typeout{** WARNING: IEEEtran.bst: No hyphenation pattern has been}%
\typeout{** loaded for the language `#1'. Using the pattern for}%
\typeout{** the default language instead.}%
\else
\language=\csname l@#1\endcsname
\fi
#2}}
\providecommand{\BIBdecl}{\relax}
\BIBdecl

\bibitem{GassendThesis}
B.~Gassend, ``Physical random functions,'' Master's thesis, M.I.T., Cambridge,
  MA, Jan. 2003.

\bibitem{WTC}
A.~D. Wyner, ``The wire-tap channel,'' \emph{Bell Labs Tech. J.}, vol.~54,
  no.~8, pp. 1355--1387, Oct. 1975.

\bibitem{benimdissertation}
O.~G{\"u}nl{\"u}, ``Key agreement with physical unclonable functions and
  biometric identifiers,'' Ph.D. dissertation, TU Munich, Germany, Nov. 2018,
  published by {Dr}. Hut Verlag.

\bibitem{AhlswedeCsiz}
R.~Ahlswede and I.~Csisz{\'a}r, ``Common randomness in information theory and
  cryptography - {P}art {I}: Secret sharing,'' \emph{IEEE Trans. Inf. Theory},
  vol.~39, no.~4, pp. 1121--1132, July 1993.

\bibitem{Maurer}
U.~M. Maurer, ``Secret key agreement by public discussion from common
  information,'' \emph{{IEEE} Trans. Inf. Theory}, vol.~39, no.~3, pp.
  2733--2742, May 1993.

\bibitem{IgnaTrans}
T.~Ignatenko and F.~M.~J. Willems, ``Biometric systems: Privacy and secrecy
  aspects,'' \emph{{IEEE} Trans. Inf. Forensics Security}, vol.~4, no.~4, pp.
  956--973, Dec. 2009.

\bibitem{LaiTrans}
L.~Lai, S.-W. Ho, and H.~V. Poor, ``Privacy-security trade-offs in biometric
  security systems - {P}art {I}: {S}ingle use case,'' \emph{{IEEE} Trans. Inf.
  Forensics Security}, vol.~6, no.~1, pp. 122--139, Mar. 2011.

\bibitem{csiszarnarayan}
I.~Csisz{\'a}r and P.~Narayan, ``Common randomness and secret key generation
  with a helper,'' \emph{IEEE Trans. Inf. Theory}, vol.~46, no.~2, pp.
  344--366, Mar. 2000.

\bibitem{bizimWZ}
O.~G{\"u}nl{\"u}, O.~\.{I}\c{s}can, V.~Sidorenko, and G.~Kramer, ``Code
  constructions for physical unclonable functions and biometric secrecy
  systems,'' \emph{{IEEE} Trans. Inf. Forensics Security}, vol.~14, no.~11, pp.
  2848--2858, Nov. 2019.

\bibitem{bizimMMMMTIFS}
O.~G{\"u}nl{\"u} and G.~Kramer, ``Privacy, secrecy, and storage with multiple
  noisy measurements of identifiers,'' \emph{{IEEE} Trans. Inf. Forensics
  Security}, vol.~13, no.~11, pp. 2872--2883, Nov. 2018.

\bibitem{bizimITW}
O.~G{\"u}nl{\"u}, R.~F. Schaefer, and G.~Kramer, ``Private authentication with
  physical identifiers through broadcast channel measurements,'' in
  \emph{{IEEE} Inf. Theory Workshop}, Visby, Sweden, Aug. 2019, pp. 1--5.

\bibitem{MerliROCorrelated}
D.~Merli, F.~Stumpf, and C.~Eckert, ``Improving the quality of ring oscillator
  {PUFs} on {FPGAs},'' in \emph{{ACM} Workshop Embedded Sys. Security}, New
  York, NY, Oct. 2010, pp. 9:1--9:9.

\bibitem{permuter}
H.~Permuter and T.~Weissman, ``Source coding with a side information
  ``{V}ending {M}achine'','' \emph{{IEEE} Trans. Inf. Theory}, vol.~57, no.~7,
  pp. 4530--4544, July 2011.

\bibitem{bizimKittipongTIFS}
O.~G{\"u}nl{\"u}, K.~Kittichokechai, R.~F. Schaefer, and G.~Caire,
  ``Controllable identifier measurements for private authentication with secret
  keys,'' \emph{{IEEE} Trans. Inf. Forensics Security}, vol.~13, no.~8, pp.
  1945--1959, Aug. 2018.

\bibitem{SecrecyviaSourcesandChannels}
V.~M. Prabhakaran, K.~Eswaran, and K.~Ramchandran, ``Secrecy via sources and
  channels,'' \emph{{IEEE} Trans. Inf. Theory}, vol.~58, no.~11, pp.
  6747--6765, Nov. 2012.

\bibitem{Blochpaper}
R.~A. Chou and M.~R. Bloch, ``Separation of reliability and secrecy in
  rate-limited secret-key generation,'' \emph{{IEEE} Trans. Inf. Theory},
  vol.~60, no.~8, pp. 4941--4957, Aug. 2014.

\bibitem{Khisti}
A.~Khisti, S.~N. Diggavi, and G.~W. Wornell, ``Secret-key generation using
  correlated sources and channels,'' \emph{{IEEE} Trans. Inf. Theory}, vol.~58,
  no.~2, pp. 652--670, Feb. 2012.

\bibitem{AminOurKeyAgreementArxiv}
A.~Gohari, O.~G{\"u}nl{\"u}, and G.~Kramer, ``Coding for positive rate in the
  source model key agreement problem,'' May 2019, [Online]. Available:
  arxiv.org/pdf/1709.05174.pdf.

\bibitem{HimanshuShun}
H.~Tyagi and S.~Watanabe, ``Converses for secret key agreement and secure
  computing,'' \emph{{IEEE} Trans. Inf. Theory}, vol.~61, no.~9, pp.
  4809--4827, Sep. 2015.

\bibitem{PappuThesis}
R.~Pappu, ``Physical one-way functions,'' Ph.D. dissertation, M.I.T.,
  Cambridge, MA, Oct. 2001.

\bibitem{bizimMDPI}
O.~G{\"u}nl{\"u}, T.~Kernetzky, O.~\.{I}\c{s}can, V.~Sidorenko, G.~Kramer, and
  R.~F. Schaefer, ``Secure and reliable key agreement with physical unclonable
  functions,'' \emph{Entropy}, vol.~20, no.~5, May 2018.

\bibitem{Transformbio}
J.~Wayman, A.~Jain, D.~Maltoni, and D.~M. (Eds), \emph{Biometric Systems:
  Technology, Design and Performance Evaluation}.\hskip 1em plus 0.5em minus
  0.4em\relax London, {U.K.}: Springer {V}erlag, Feb. 2005.

\bibitem{MatthieuPolar}
R.~A. Chou, M.~R. Bloch, and E.~Abbe, ``Polar coding for secret-key
  generation,'' \emph{{IEEE} Trans. Inf. Theory}, vol.~61, no.~11, pp.
  6213--6237, Nov. 2015.

\bibitem{OSRBAmin}
M.~H. Yassaee, M.~R. Aref, and A.~Gohari, ``Achievability proof via output
  statistics of random binning,'' \emph{{IEEE} Trans. Inf. Theory}, vol.~60,
  no.~11, pp. 6760--6786, Nov. 2014.

\bibitem{RoyCondLN}
R.~Timo, T.~J. Oechtering, and M.~Wigger, ``Source coding problems with
  conditionally less noisy side information,'' \emph{{IEEE} Trans. Inf.
  Theory}, vol.~60, no.~9, pp. 5516--5532, Sep. 2014.

\bibitem{SRAMPUFs}
R.~Maes, P.~Tuyls, and I.~Verbauwhede, ``A soft decision helper data algorithm
  for {SRAM PUFs},'' in \emph{{IEEE} Int. Symp. Inf. Theory}, Seoul, South
  Korea, June 2009, pp. 2101--2105.

\bibitem{bizimpaper}
O.~G{\"u}nl{\"u} and O.~\.{I}\c{s}can, ``{DCT} based ring oscillator physical
  unclonable functions,'' in \emph{{IEEE} Int. Conf. Acoust., Speech Sign.
  Proc.}, Florence, Italy, May 2014, pp. 8198--8201.

\bibitem{bizimtemperaturepaper}
O.~G{\"u}nl{\"u}, O.~\.{I}\c{s}can, and G.~Kramer, ``Reliable secret key
  generation from physical unclonable functions under varying environmental
  conditions,'' in \emph{{IEEE} Int. Workshop Inf. Forensics Security}, Rome,
  Italy, Nov. 2015, pp. 1--6.

\bibitem{BlochLectureNotes2018}
M.~Bloch, \emph{{L}ecture {N}otes in {I}nformation-{T}heoretic
  {S}ecurity}.\hskip 1em plus 0.5em minus 0.4em\relax Atlanta, GA: Georgia
  Inst. Technol., July 2018.

\bibitem{SW}
D.~Slepian and J.~Wolf, ``Noiseless coding of correlated information sources,''
  \emph{{IEEE} Trans. Inf. Theory}, vol.~19, no.~4, pp. 471--480, July 1973.

\bibitem{Elgamalbook}
A.~E. Gamal and Y.-H. Kim, \emph{Network {I}nformation {T}heory}.\hskip 1em
  plus 0.5em minus 0.4em\relax Cambridge, {U.K.}: Cambridge {U}niversity
  {P}ress, 2011.

\bibitem{Blochbook}
M.~Bloch and J.~Barros, \emph{Physical-layer {S}ecurity}.\hskip 1em plus 0.5em
  minus 0.4em\relax Cambridge, {U.K.}: Cambridge {U}niversity {P}ress, 2011.

\bibitem{ChandraLessNoisy}
Z.~V. Wang and C.~Nair, ``The capacity region of a class of broadcast channels
  with a sequence of less noisy receivers,'' in \emph{{IEEE} Int. Symp. Inf.
  Theory}, Austin, TX, June 2010, pp. 595--598.

\bibitem{CsiszarKornerbook2011}
I.~Csisz{\'a}r and J.~K{\"o}rner, \emph{Information {T}heory: {C}oding
  {T}heorems for {D}iscrete {M}emoryless {S}ystems}, 2nd~ed.\hskip 1em plus
  0.5em minus 0.4em\relax Cambridge, {U.K}.: Cambridge {U}niversity {P}ress,
  2011.

\end{thebibliography}

\end{document}